# An Adynamical, Graphical Approach to Quantum Gravity and Unification

*W.M. Stuckey[1], Michael Silberstein[2,3] and Timothy McDevitt[4]*

## Abstract

We use graphical field gradients in an adynamical, background independent fashion to propose a new approach to quantum gravity and unification. Our proposed reconciliation of general relativity and quantum field theory is based on a modification of their graphical instantiations, i.e., Regge calculus and lattice gauge theory, respectively, which we assume are fundamental to their continuum counterparts. Accordingly, the fundamental structure is a graphical amalgam of space, time and sources (in parlance of quantum field theory) called a "spacetimesource element." These are fundamental elements *of* space, time, and sources, not source elements *in* space and time. The transition amplitude for a spacetimesource element is computed using a path integral with discrete graphical action. The action for a spacetimesource element is constructed from a difference matrix K and source vector J on the graph, as in lattice gauge theory. K is constructed from graphical field gradients so that it contains a non-trivial null space and J is then restricted to the row space of K, so that it is divergence-free and represents a conserved exchange of energy-momentum. This construct of K and J represents an adynamical global constraint between sources, the spacetime metric, and the energy-momentum content of the element, rather than a dynamical law for time-evolved entities. In this view, one manifestation of quantum gravity becomes evident when, for example, a single spacetimesource element spans adjoining simplices of the Regge calculus graph. Thus energy conservation for the spacetimesource element includes contributions to the deficit angles between simplices. This idea is used to correct proper distance in the Einstein-deSitter cosmology model yielding a fit of the Union2 Compilation supernova data that matches ΛCDM without having to invoke accelerating expansion or dark energy. A similar modification to lattice gauge theory results in an adynamical account of quantum interference.

[1] Department of Physics, Elizabethtown College, Elizabethtown, PA 17022, stuckeym@etown.edu
[2] Department of Philosophy, Elizabethtown College, Elizabethtown, PA 17022, silbermd@etown.edu
[3] Department of Philosophy, University of Maryland, College Park, MD 20742
[4] Department of Mathematics, Elizabethtown College, Elizabethtown, PA 17022, mcdevittt@etown.edu

## 1. Introduction

*1.1 Overview*. In this paper, we introduce our relational, adynamical, background independent approach to quantum gravity (QG) and the unification of physics. This approach is based on and motivated by our foundations-driven account of quantum physics called *Relational Blockworld*(1) (RBW[1]), and employs methods from general relativity (background independence and variable geometry), particle physics (path integral formalism), and lattice gauge theory (graphical construction of transition amplitude). More specifically, we propose a reconciliation of general relativity (GR) and quantum field theory (QFT) via modification of their graphical instantiations, i.e., Regge calculus (section 5.1) and lattice gauge theory (LGT), respectively, which we assume are fundamental to their continuum counterparts. The modifications we propose deal with our fundamental ontological elements of quantum physics, i.e., graphical amalgams of space, time and sources[2] that we call "spacetimesource elements" (Figure 1). Accordingly, these are elements *of* space, time, and sources, not source elements *in* space and time. The source of a spacetimesource element is divergence-free and represents an unmediated, conserved exchange of energy-momentum. Our approach constitutes a modification of LGT and Regge calculus in three respects.

First, we're assuming QFT and GR are continuum approximations of LGT and Regge calculus, respectively, which is the opposite of conventional thinking. Second, we are underwriting the fundamental computational element of LGT, i.e., the transition amplitude, in a relational and adynamical fashion. More significantly, third, we are assuming that the size of spacetimesource elements of LGT and simplices of Regge calculus (Figure 13) can be as small or large as the situation requires. We find these changes discharge the technical and conceptual difficulties of QFT and quantum mechanics (QM) while leaving their computational structures and empirical successes intact, for all practical purposes(2). For example, the flexibility in element/simplex size provides an adynamical explanation of twin-slit interference (section 3.4) and a novel solution to the dark energy problem (section 5.2). We expect of course that this view of fundamental physics will suggest new experiments in other areas as well. We will only briefly touch on such issues here, but some consequences will be obvious to the reader familiar with quantum physics. The focus of this paper will be on explaining the mathematical and conceptual structure of the spacetimesource element of our "modified lattice gauge theory" (MLGT) in the Schrödinger, Klein-Gordon, Dirac, Maxwell, and Einstein-Hilbert actions, with extension to the Standard Model of particle physics[3] and consequences for QG and unification.

[1] RBW as a realist psi-epistemic retrocausal interpretation of quantum mechanics is published elsewhere http://www.ijqf.org/archives/2087, so this paper will focus on our "modified lattice gauge theory" approach to quantum gravity and unification.

[2] We use the word "source" in formal analogy to quantum field theory (QFT) where it means "particle sources" or "particle sinks" (creation or annihilation events, respectively). However, our "sources" are not always equivalent to the sources in QFT, just as our "fields" are not always equivalent to those of quantum physics. When we want to specify "a source of energy-momentum" we will use "Source." As we will explain in mathematical detail in section 3.2, our "source" represents a divergence-free property of the spacetimesource element determined contextually by "classical objects," i.e., objects with a worldline/tube in spacetime.

We finish section 1 with an overview of our proposed formalism, a discussion of locality issues associated with our beable (the spacetimesource element), and our proposed method of adynamical explanation. In section 2, we articulate more of our ontology, i.e., ontic structural realism in a block universe. Section 3 contains formal details associated with the construct of the spacetimesource element in the Schrödinger, Klein-Gordon, Dirac, Maxwell, and Einstein-Hilbert actions. The spacetimesource element of the Schrödinger action is used to provide an adynamical explanation of twin-slit interference. The approach is extended to the spacetimesource elements for the Standard Model in section 4, where we also explain our view of particle physics to include UV and IR regularization. In section 5, we outline consequences for astrophysics and cosmology, to include an overview of our resolution of the dark energy problem via modified Regge calculus (MORC). We find that correcting proper distance in the Einstein-deSitter (EdS) cosmology model according to MORC yields a fit(3) of the Union2 Compilation supernova data that matches ΛCDM without having to invoke accelerating expansion or dark energy(4). We conclude in section 6 with a list of outstanding questions for our program and consequences for unification and QG.

---

[3] Hereafter simply "the Standard Model."

In the language of dynamism, our spacetimesource elements represent quantum exchanges of energy-momentum (in the form of mass, charge, or other property) between objects with worldlines/tubes[4] in spacetime, e.g., Sources, beam splitters, mirrors, detectors, etc. Let us call such objects with worldlines/tubes in spacetime "classical objects" (COs). In a quantum exchange between COs per RBW, energy-momentum disappears from one CO and reappears later in another (Figure 1), i.e., the quantum exchange is unmediated. The energy-momentum loss and gain are two distinct events involving two distinct COs separated in space and time where the spatial separation, temporal duration, and energy-momentum exchanged are all associated with one spacetimesource element. Obviously, mediated exchanges can occur between COs as well. For example, a buckyball that decoheres between Source, diffraction grating, and detector such that it does not contribute to an interference pattern at the detector could be represented by a worldline. Such a buckyball would therefore constitute a CO mediating a (non-quantum) exchange of mass between Source and detector. However, a buckyball Source and diffraction grating creating an interference pattern at the detector[(5)] would represent a set of quantum exchanges which would not have worldlines (section 3.4). This is in stark contrast to a view like deBroglie-Bohm where a buckyball interference pattern would be created by objects with wordlines guided by waves. As we have written elsewhere[(6)], a dynamical, algebraic counterpart to the spacetimesource element in the path integral of RBW might be "elementary process" in the Clifford algebra of Hiley's Implicate Order.

The notion that energy might disappear from one location and reappear at another without traversing the space between has also been claimed by the authors of the DFBV experiment[(7)], Danan, Farfurnik, Bar-Ad, and Vaidman[5]. Their conclusion is based on so-called "weak measurements" of photons in an interferometer that contains a "nested interferometer." They explain their weak measurement results assuming photons exist inside the nested interferometer even though those photons "never entered and never left the nested interferometer." They therefore conclude that, "The photons do not always follow continuous trajectories," just as we assert concerning quantum exchange in general[8].

---

[4] We follow convention and use "worldtubes" for spatially extended objects and "worldlines" for objects treated as points.

[5] For another so-called "direct-action" approach to quantum gravity see Wesley, D., & Wheeler, J. A.: Towards an action-at-a-distance concept of spacetime: In: Ashtekar, A., *et al.* (eds.) Revisiting the Foundations of Relativistic Physics: Festschrift in Honor of John Stachel. Boston Studies in the Philosophy and History of Science (Book 234), pp 421-436, Kluwer Academic Publishers, Dordrecht, the Netherlands (2003).

So just to be clear, in the parlance of RBW, the properties associated with spacetimesource elements – spatial and temporal extent, mass, charge, etc. – are determined relationally between COs, they are not intrinsic properties *of* individual spacetimesource elements. That is, the existence of these properties requires a collection of connected spacetimesource elements in the context of COs, e.g., mass and charge (properties) associated with tracks (connected spacetimesource elements) in a particle detector (CO) as we explain in section 4.2. Essentially, we are claiming that the worldtube of any particular CO *in* space and time (defined relationally by its surrounding COs) can be decomposed into spacetimesource elements *of* space, time, and sources organized per an adynamical global constraint (AGC) using the context of those surrounding COs. Herein we articulate the part of that constraint dealing with individual spacetimesource elements.

To do so, we relate gauge invariance, gauge fixing, divergence-free sources, and relationally defined interacting COs in an adynamic, graphical fashion. Specifically, each row of our difference matrix $\vec{\vec{K}}$ for field gradients in the graphical action for our spacetimesource element is a vector constructed relationally via the connectivity of some graphical entity, i.e., nodes connected by links, links connected by plaquettes, or plaquettes connected by cubes. Since each row vector is relationally defined, its components sum to zero, which means $[111\ldots]^T$ is a null eigenvector of $\vec{\vec{K}}$. Our AGC then demands that the source vector $\vec{J}$ in the action for our spacetimesource element reside in the row space of $\vec{\vec{K}}$, so that it is orthogonal to $[111\ldots]^T$ which means its components sum to zero, i.e., it is divergence-free. A divergence-free source in each spacetimesource element then underwrites relationally defined, spatially distributed, trans-temporally identified (conserved) properties exchanged between interacting COs, i.e., it provides the fundamental ontological element for relationally/contextually defined COs per ontic structural realism (section 2). That $\vec{\vec{K}}$ possesses a non-trivial null space is the graphical equivalent of gauge invariance and restricting $\vec{J}$ to the row space of $\vec{\vec{K}}$ provides a natural gauge fixing, i.e., restricting the path integral of the transition amplitude to the row space of $\vec{\vec{K}}$. That $\vec{\vec{K}}$ possesses a non-trivial null space also means the determinant of $\vec{\vec{K}}$ is zero, so the set of vectors constituting the rows of $\vec{\vec{K}}$ is not linearly independent. That some subset of these row vectors is determined by its complement follows from having the graphical set relationally constructed. Thus, divergence-free $\vec{J}$ follows from relationally defined $\vec{\vec{K}}$ as a direct result of our AGC. Consequently, we agree with Rovelli that[9], "Gauge is ubiquitous. It is not unphysical redundancy of our mathematics. It reveals the relational structure of our world."

So these fundamental elements of spacetimesource are our beables. Are such beables local?

*1.2 Locality*. Concerning the locality of beables, Einstein writes[10]

> ..if one asks what is characteristic of the realm of physical ideas independently of the quantum theory, then above all the following attracts our attention: the concepts of physics refer to a real external world, i.e., ideas are posited of things that claim a 'real existence' independent of the perceiving subject (bodies, fields, etc.), and these ideas are, on the other hand, brought into as secure a relationship as possible with sense impressions. Moreover, it is characteristic of these physical things that they are conceived of as being arranged in a spacetime continuum. Further, it appears to be essential for this arrangement of the things introduced in physics that, at a specific time, these things claim an existence independent of one another, insofar as these things 'lie in different parts of space'. Without such an assumption of mutually independent existence (the 'being-thus') of spatially distant things, an assumption which originates in everyday thought, physical thought in the sense familiar to us would not be possible....

There has been a great deal of handwringing lately in the foundations literature on QG as to whether the most fundamental unifying theory from which spacetime emerges, must have local beables to be empirically coherent and make full correspondence with higher-level physical theories and the experienced world[11]. Maudlin notes that[12] "local beables do not merely exist: they exist somewhere," or as Bell puts it[13], beables are "definitely associated with particular space-time regions." Of course there is less consensus about the necessary and sufficient conditions for being a local beable, and that discussion is beyond the scope of this paper. In any case, we share the consensus view that a successful theory of quantum gravity need not have local beables[14].To return to the main question about the status of spacetimesource elements, local beables are thought of as being separate from but located somewhere *in* spacetime, whereas, again, spacetimesources are *of* space, time and sources.

Einstein appears to conflate (or at least highlight) several different notions of "local" in the passage above, including, (1) local as localized in spacetime, (2) local as possessing primitive thisness with intrinsic properties, (3) local as in no superluminal interactions and (4) local as in being otherwise independent (e.g., statistically) of entities at other points in spacetime. On the graph, directly connected nodes are "local" to each other. However, due to the third modification of LGT and Regge calculus described above, it can be the case (see section 5.2) that neighboring points on the graph are very far apart in the spacetime manifold M, so those points are not local/neighboring in M (this is called "disordered locality"). Thus, our beables are local in the first sense per graphical locality, and local in the third sense on both the graph and its continuum approximation, the spacetime manifold M (see section 5.1). Again, recalling what we said above, our beables are not local in senses 2 or 4.

The manner by which we correct EdS cosmology and explain quantum interference is a form of disordered locality, as explained supra, similar to the situation in quantum graphity[15]. Our physical model thus implements a suggestion made by Weinstein among others[16]:

> What I want to do here is raise the possibility that there is a more fundamental theory possessing nonlocal constraints that underlies our current theories. Such a theory might account for the mysterious nonlocal effects currently described, but not explained, by quantum mechanics, and might additionally reduce the extent to which cosmological models depend on finely tuned initial data to explain the large scale correlations we observe. The assumption that spatially separated

> physical systems are entirely uncorrelated is a parochial assumption borne of our experience with the everyday objects described by classical mechanics. Why not suppose that at certain scales or certain epochs, this independence emerges from what is otherwise a highly structured, nonlocally correlated microphysics?

As he says, every extant fundamental theory of physics assumes the non-existence of such nonlocal constraints[(17)]:

> Despite radical differences in their conceptions of space, time, and the nature of matter, all of the physical theories we presently use, non-relativistic and relativistic, classical and quantum, share one assumption: the features of the world at distinct points in space are understood to be independent. Particles may exist anywhere, independent of the location or velocity of other particles. Classical fields may take on any value at a given point, constrained only by local constraints like Gauss's law. Quantum field theories incorporate the same independence in their demand that field operators at distinct points in space commute with one another. The independence of physical properties at distinct points is a theoretical assumption, albeit one that is grounded in our everyday experience. We appear to be able to manipulate the contents of a given region of space unrestricted by the contents of other regions. We can arrange the desk in our office without concern for the location of the couch at home in our living room.

RBW provides an exact model in which precisely this type of locality (type 2 and type 4 above) fails to obtain, thereby allowing us to explain a diverse range of phenomena from quantum interference to so-called dark energy. Furthermore, as will be explained, the failure of locality in question, the way is it implemented in our theory, is consistent with and driven by an appropriately modified Regge calculus.

*1.3 Adynamical Explanation*. Our approach also differs from common practice (even quantum graphity) in that it is *adynamical*[(18)]. Carroll sums up nicely what we mean by a dynamical approach[(19)]:

> Let's talk about the actual way physics works, as we understand it. Ever since Newton, the paradigm for fundamental physics has been the same, and includes three pieces. First, there is the "space of states": basically, a list of all the possible configurations the universe could conceivably be in. Second, there is some particular state representing the universe at some time, typically taken to be the present. Third, there is some rule for saying how the universe evolves with time. You give me the universe now, the laws of physics say what it will become in the future. This way of thinking is just as true for quantum mechanics or general

> relativity or quantum field theory as it was for Newtonian mechanics or Maxwell's electrodynamics.

Carroll goes on to say that all extant formal models of QG, even those attempting to recover spacetime[20], are dynamical in this sense. While it is true that integral calculus and least action principles have been around for a long time, most assume these methods are formal tricks and not fundamental to dynamical equations. While our adynamical approach employs mathematical formalism akin to dynamical theories, e.g., LGT, we redefine what it means to "explain" something in physics. Rather than finding a rule for time-evolved entities per Carroll (e.g., causal dynamical triangulation[21]), the AGC leads to the self-consistency of a graphical spacetime metric and its relationally defined sources. While we do talk about "constructing" or "building" spatiotemporal objects in this paper, we are not implying any sort of "evolving block universe" as in causet dynamics[22]. Our use of this terminology is merely in the context of a computational algorithm. So, one might ask for example, "Why does link X have metric *G* and stress-energy tensor T?" A dynamical answer might be, "Because link X-1 has metric *G* -1 and stress-energy tensor T-1 and the law of evolution thereby dictates that link X has metric *G* and stress-energy tensor T." Notice how this answer is independent of future boundary conditions; indeed, it's independent of conditions anywhere else on the graph other than those of the 3D hypersurface in the immediate past. Contrast this with an adynamical answer such as, "Because the values *G* and T on X satisfy the AGC for the graph as a whole, given input anywhere in the past, present, and/or future of X." For example, we will argue in section 4.2 that particle trajectories of high energy physics experiments satisfy the AGC given the spacetime configuration of colliding beams and detector. It should be clear from what we have said thus far that explanations involving adynamical global constraints typically involve future boundary conditions, which brings us to the next section.

## 2. Quantum Physics Reconceived: Ontic Structural Realism in a Block Universe

*2.1 Dynamism Denied.* Our account of spacetime and matter is very much in keeping with Rovelli's intuition that[23]:

> General relativity (GR) altered the classical understanding of the concepts of space and time in a way which...is far from being fully understood yet. QM

> challenged the classical account of matter and causality, to a degree which is still the subject of controversies. After the discovery of GR we are no longer sure of what is space-time and after the discovery of QM we are no longer sure of what matter is. *The very distinction between space-time and matter is likely to be ill-founded*....I think it is fair to say that today we do not have a consistent picture of the physical world. [italics added]

We agree with Rovelli and believe a current obstacle to unification is the lack of a true marriage of spacetime with matter. That is, we believe one of the main obstacles to unification has been a form of 'spacetime-matter dualism' whereby the spacetime metric (or simply "metric") is subject to quantization distinct from the matter and gauge fields. This view is carried over from QFT and GR. In QFT, although matter-energy fields are imagined to pervade space, the metric is independent of the matter-energy content of spacetime. And, although Weyl characterized GR as providing *RaumZeitMaterie*(24), there are vacuum solutions in GR, i.e., spacetime regions where the stress-energy tensor is zero. Thus, neither QFT nor GR embody a true unity of "spacetimesource" and both employ a differentiable manifold structure for spacetime[6]. Herein we propose unification based on a true unity of space, time, and sources, finishing Einstein's dream so to speak.

Fundamental theories of physics (e.g., M-theory, loop quantum gravity, causets) may deviate from the norm by employing radical new fundamental entities (e.g., branes, loops, ordered sets), but the game is always dynamical, broadly construed (e.g., vibrating branes, geometrodynamics, sequential growth process). As Healey puts it(25):

> Physics proceeds by first analyzing the phenomena with which it deals into various kinds of systems, and then ascribing states to such systems. To classify an object as a certain kind of physical system is to ascribe certain, relatively stable, qualitative intrinsic properties; and to further specify the state of a physical system is to ascribe to it additional, more transitory [time dependent], qualitative intrinsic properties….A physical property of an object will then be both qualitative and intrinsic just in case its possession by that object is wholly determined by the underlying physical states and physical relations of all the basic systems that compose that object.

Dynamism then encompasses three claims: (A) the world, just as appearances and the experience of time suggest, evolves or changes in time in some objective fashion, (B) the best explanation for A will be some dynamical law that "governs" the evolution of the system in question, and (C) the fundamental entities in a "theory of everything" will themselves be dynamical entities evolving in some space however abstract, e.g., Hilbert space. Our model rejects not only tenets A and B of dynamism, but also C. In our view, time-evolved *entities* or *things* are not fundamental and, in fact, it is in accord with ontic structural realism(26) (OSR):

> Ontic structural realists argue that what we have learned from contemporary physics is that the nature of space, time and matter are not compatible with standard metaphysical views about the ontological relationship between individuals, intrinsic properties and relations. On the broadest construal OSR is any form of structural realism based on an ontological or metaphysical thesis that inflates the ontological priority of structure and relations.

---

[6] For an overview of problems associated with "the manifold conception of space and time" in quantum gravity see Butterfield, J., & Isham, C.J.: Spacetime and the Philosophical Challenge of Quantum Gravity (1999) http://arxiv.org/abs/gr-qc/9903072.

More specifically, our RBW version of OSR agrees that(27) "The relata of a given relation always turn out to be relational structures themselves on further analysis." Note that OSR does not claim there are relations without relata, just that the relata are not individuals (e.g., things with primitive thisness and intrinsic properties), but always ultimately analyzable as relations as well (Figure 1). OSR already somewhat violates the dynamical bias by rejecting *things* with intrinsic properties as fundamental *building blocks* of reality – the world isn't fundamentally *compositional* – the deepest conception of reality is not one in which we decompose things into other things at ever smaller length and time scales[7]. Our beables (spacetimesource elements) are certainly a violation of a compositional picture of realty, since their properties are inherited from their classical context. We however go even further in rejecting dynamism, not merely because it is a block universe, but because the fundamental modal structure, the fundamental AGC, is not a dynamical law or even spacetime symmetries.

A good deal of the literature on OSR is driven by philosophical concerns about scientific realism and intertheoretic relations, rather than motivated by physics itself(28). There has also been much debate in the philosophical literature as to whether OSR provides any real help in resolving foundational issues of physics such as interpreting QM or in advancing physics itself. Consider the following claims for example:

> OSR is not an interpretation of QM in addition to many worlds-type interpretations, collapse-type interpretations, or hidden variable-type interpretations. As the discussion of the arguments for OSR from QM in section 2 above has shown, OSR is not in the position to provide on its own an ontology for QM, since it does not reply to the question of what implements the structures that it poses. In conclusion, after more than a decade of elaboration and debate on OSR about QM, it seems that the impact that OSR can have on providing an answer to the question of what the world is like, if QM is correct, is rather limited. From a scientific realist perspective, the crucial issue is the assessment of the pros and cons of the various detailed proposals for an ontology of QM, as it was before the appearance of OSR on the scene(29).

> While the basic idea defended here (a fundamental ontology of brute relations) can be found elsewhere in the philosophical literature on 'structural realism', we have yet to see the idea used as an argument for advancing physics, nor have we seen a truly convincing argument, involving a real construction based in modern physics, that successfully evades the objection that there can be no relations without first (in logical order) having things so related(30).

[7] This is an ontological claim. Computationally, of course, the spacetime lattice of LGT is "composed of" hypercubes with fields on nodes and links.

As this paper will attest, RBW is a counterexample to Esfeld's claim and it provides exactly the physical model that Rickles & Bloom are looking for. As they say in the following passage, OSR has the potential to re-ground physics, dissolve current quagmires and lead to new physics(31):

> Viewing the world as structurally constituted by primitive relations has the potential to lead to new kinds of research in physics, and knowledge of a more stable sort. Indeed, in the past those theories that have adopted a broadly similar approach (along the lines of what Einstein labeled 'principle theories') have led to just the kinds of advances that this essay competition seeks to capture: areas "where thinkers were 'stuck' and had to let go of some cherished assumptions to make progress." Principle theory approaches often look to general 'structural aspects' of physical behaviour over 'thing aspects' (what Einstein labeled 'constructive'), promoting invariances of world-structure to general principles.

Rickles & Bloom lament the fact that OSR has yet to be so motivated and further anticipate our theory almost perfectly when they say(32):

> The position I have described involves the idea that physical systems (which I take to be characterized by the values for their observables) are exhausted by extrinsic or relational properties: they have no intrinsic, local properties at all! This is a curious consequence of background independence coupled with gauge invariance and leads to a rather odd picture in which objects and [spacetime] structure are deeply entangled. Inasmuch as there are objects at all, any properties they possess are structurally conferred: they have no reality outside some correlation. What this means is that the objects don't *ground* structure, they are nothing independently of the structure, which takes the form of a (gauge invariant) correlation between (non-gauge invariant) field values. With this view one can both evade the standard 'no relations without relata' objection and the problem of accounting for the appearance of time (in a timeless structure) in the same way.

In this paper we provide physics that embodies their suggestion.

*2.2 Block Universe*. As stated, we must further exacerbate this violation of dynamism by applying OSR to a block universe. The block universe perspective (the reality of all events past, present, and future including the outcomes of quantum experiments) is suggested for example by the relativity of simultaneity in special relativity or, more generally, the lack of a preferred spatial foliation of spacetime M in GR, and even by quantum entanglement according to some of us[(33)]. Geroch writes[(34)]:

> There is no dynamics within space-time itself: nothing ever moves therein; nothing happens; nothing changes. In particular, one does not think of particles as moving through space-time, or as following along their world-lines. Rather, particles are just in space-time, once and for all, and the world-line represents, all at once, the complete life history of the particle.

When Geroch says that "there is no dynamics within space-time itself," he is not denying that the mosaic of the block universe possesses patterns that can be described with dynamical laws. Nor is he denying the predictive and explanatory value of such laws. Rather, given the reality of all events in a block universe, dynamics are not "event factories" that bring heretofore non-existent events (such as measurement outcomes) into being; fundamental dynamical laws that are allegedly responsible for discharging fundamental "why" questions in physics are not brute unexplained explainers that "produce" events on our view.

In addition there is the problem of time in canonical general relativity. That is, in a particular Hamiltonian formulation of GR the reparametrization of spacetime is a gauge symmetry. Therefore, all genuinely physical magnitudes are constants of motion, i.e., they don't change over time. In short, as Rovelli stated in an earlier quote, gauge invariance merely "reveals the relational structure of our world."

Finally, the problem of frozen time in canonical QG is that if the canonical variables of the theory to be quantized transform as scalars under time reparametrizations, which is true in practice because they have a simple geometrical meaning, then(35) "the Hamiltonian is (weakly) zero for a generally covariant system." The result upon canonical quantization is the famous Wheeler-DeWitt equation, void of time evolution. While it is too strong to say a generally covariant theory must have $H = 0$, there is no well-developed theory of quantum gravity that has avoided it to date(36). It is supremely ironic that the dynamism and unificationism historically driving physics led us directly to block universe and frozen time, but RBW discharges the irony.

Just as people rarely take seriously the possibility that the path integral or Lagrangian approach (with its future boundary conditions) is fundamental(37), they rarely take seriously the block universe even when they embrace it. That is, the overriding assumption as noted above is that dynamical explanation is fundamental. But taking the block universe seriously opens up the possibility that adynamical global constraints might be fundamental both for individual systems within the universe, e.g., the delayed-choice experiment, and for the universe as a whole, e.g., the big bang (the allegedly special and unlikely initial conditions).

Back to the problem of time, Rickles notes that the problem can be solved by(38), "(1) global quantities defined over the whole spacetime and (2) 'relational' quantities built out of correlations between field values and/or invariants. There seems to be some consensus forming that the latter type are the way to go, and these will serve as the appropriate vehicle for defining time in an unchanging mathematical structure, as well as defining the structures themselves." Our theory, it will become clear, provides a solution precisely in terms of number 2.

We think therefore that both QM, e.g., delayed-choice experiments, and relativity are telling us that Nature is a block universe, so it is time to promote this idea from mere metaphysics to physics. This is what RBW does.

*2.3 OSR in a Block Universe*. Putting it all together, reality is a block universe best characterized as spacetimesource, as opposed to the "spacetime + sources" picture of current physics. In the foundations literature on the eternalism debate and the structural realism debate respectively, the biggest complaint is that the fate of these topics makes no real difference for physics itself, i.e., it does not lead to new models, new insights, or new predictions and it does not resolve conceptual problems. In short, the complaint is that such debates are nothing but pure metaphysics. We, however, actually do provide a new formal model for fundamental physics based on the block universe with relationally defined sources that has all the aforementioned virtues. Our approach employs an adynamical global constraint.

*2.4 Adynamical Global Constraint*. Our use of an AGC is not without precedent, as we already have an ideal example in Einstein's equations of GR

$$R_{\alpha\beta} - \frac{1}{2} g_{\alpha\beta} R = \frac{8\pi G}{c^4} T_{\alpha\beta}$$

Momentum, force, and energy all depend on spatiotemporal measurements (tacit or explicit), so the stress-energy tensor $T_{\alpha\beta}$ cannot be constructed without tacit or explicit knowledge of the spacetime metric $g_{\alpha\beta}$ (technically, $T_{\alpha\beta}$ can be written as the functional derivative of the matter-energy Lagrangian with respect to $g^{\alpha\beta}$). But, if one wants a "dynamic spacetime" in the parlance of GR, $g_{\alpha\beta}$ must depend on the matter-energy distribution in spacetime. GR solves this dilemma by constraining $T_{\alpha\beta}$ to be "consistent" with $g_{\alpha\beta}$ everywhere on the spacetime manifold M per Einstein's equations[8]. This adynamical global constraint hinges on divergence-free sources, which finds a mathematical counterpart in $\partial\partial = 0$, i.e., the boundary of a boundary principle(39). So, Einstein's equations of GR are a mathematical articulation of the boundary of a boundary principle in classical physics[9], i.e., they constitute an adynamical global constraint in classical physics In fact, our AGC at the level of the spacetimesource element is based on the same topological maxim ($\partial\partial = 0$) for the same reason, as is the case with quantum and classical electromagnetism(40).

## 3. Underwriting the Free Field Transition Amplitude

*3.1 Boundary of a Boundary Principle*. In Figure 2, the boundary of plaquette $\mathbf{p}_1$ is given by links $\mathbf{e}_4 + \mathbf{e}_5 - \mathbf{e}_2 - \mathbf{e}_1$, which also provides an orientation. The boundary of $\mathbf{e}_1$ is given by nodes $\mathbf{v}_2 - \mathbf{v}_1$, which likewise provides an orientation. Using these conventions for the orientations of links and plaquettes we have the following boundary operator for $C_2 \rightarrow C_1$, i.e., space of plaquettes mapped to space of links in the spacetime chain complex:

$$\partial_2 = \begin{bmatrix} -1 & 0 \\ -1 & 1 \\ 0 & -1 \\ 1 & 0 \\ 1 & 0 \\ 0 & 1 \\ 0 & -1 \end{bmatrix} \qquad (1)$$

The first column is simply the links for the boundary of $\mathbf{p}_1$ and the second column is simply the links for the boundary of $\mathbf{p}_2$. We have the following boundary operator for $C_1 \rightarrow C_0$, i.e., space of links mapped to space of nodes in the spacetime chain complex:

---

[8] Concerning the stress-energy tensor, Hamber and Williams write, "In general its covariant divergence is not zero, but consistency of the Einstein field equations demands $\nabla^{\alpha} T_{\alpha\beta} = 0$," Hamber, H.W., & Williams, R.: Nonlocal Effective Gravitational Field Equations and the Running of Newton's G (2005) http://arxiv.org/pdf/hep-th/0507017.pdf

[9] The fundamental ontological entities of GR are described via worldlines/tubes, so it admits a dynamical interpretation, of course. The adynamical and global nature of a GR explanation is more evident in its discrete graphical counterpart, Regge calculus (section 5.1).

$$\partial_1 = \begin{bmatrix} -1 & 0 & 0 & -1 & 0 & 0 & 0 \\ 1 & -1 & -1 & 0 & 0 & 0 & 0 \\ 0 & 0 & 1 & 0 & 0 & 0 & -1 \\ 0 & 0 & 0 & 1 & -1 & 0 & 0 \\ 0 & 1 & 0 & 0 & 1 & -1 & 0 \\ 0 & 0 & 0 & 0 & 0 & 1 & 1 \end{bmatrix} \tag{2}$$

which completes the spacetime chain complex, $C_0 \xleftarrow{\partial_1} C_1 \xleftarrow{\partial_2} C_2$. The columns are simply the nodes for the boundaries of the edges or conversely, each row shows which links leave (-1) or enter (1) each node. These boundary operators satisfy $\partial_1\partial_2 = 0$ as required by the boundary of a boundary principle.

*3.2 Graphical Harmonic Oscillator and the Adynamical Global Constraint*. The Lagrangian for the coupled masses of Figure 3 is

$$L = \frac{1}{2}m\dot{q}_1^2 + \frac{1}{2}m\dot{q}_2^2 - \frac{1}{2}k(q_1 - q_2)^2 \tag{3}$$

so our transition amplitude is ($\hbar = 1$)

$$Z = \int Dq(t)\exp\left[i\int_0^T dt\left[\frac{1}{2}m\dot{q}_1^2 + \frac{1}{2}m\dot{q}_2^2 - \frac{1}{2}kq_1^2 - \frac{1}{2}kq_2^2 + kq_1q_2 + J_1q_1 + J_2q_2\right]\right] \tag{4}$$

giving

$$\vec{\vec{K}} = \begin{bmatrix} \left(\frac{m}{\Delta t} - k\Delta t\right) & \frac{-m}{\Delta t} & 0 & k\Delta t & 0 & 0 \\ \frac{-m}{\Delta t} & \left(\frac{2m}{\Delta t} - k\Delta t\right) & \frac{-m}{\Delta t} & 0 & k\Delta t & 0 \\ 0 & \frac{-m}{\Delta t} & \left(\frac{m}{\Delta t} - k\Delta t\right) & 0 & 0 & k\Delta t \\ k\Delta t & 0 & 0 & \left(\frac{m}{\Delta t} - k\Delta t\right) & \frac{-m}{\Delta t} & 0 \\ 0 & k\Delta t & 0 & \frac{-m}{\Delta t} & \left(\frac{2m}{\Delta t} - k\Delta t\right) & \frac{-m}{\Delta t} \\ 0 & 0 & k\Delta t & 0 & \frac{-m}{\Delta t} & \left(\frac{m}{\Delta t} - k\Delta t\right) \end{bmatrix} \tag{5}$$

on the graph of Figure 4. The null space (space of eigenvalues 0) is spanned by the eigenvector $[111111]^{\mathrm{T}}$. The space orthogonal to the null space of $\vec{\vec{K}}$ is called the row space[10] of $\vec{\vec{K}}$. Therefore, any source vector $\vec{J}$ in the row space of $\vec{\vec{K}}$ has components which sum to zero and this is referred to in graphical approaches to physics as "divergence-free $\vec{J}$." If $\vec{J}$ is a force, this simply reflects Newton's third law. If $\vec{J}$ is energy, this simply reflects conservation of energy. We will use $\vec{J}$ on spacetimesource elements to underwrite conserved properties defining TTOs, so we require that $\vec{J}$ reside in the row space of $\vec{\vec{K}}$, as well as represent an interaction with conserved source across a spacetimesource element. Thus, $\vec{\vec{K}}$ must be constructed so as to possess a non-trivial null space, which is the graphical equivalent of gauge invariance.

Giving weights to the links of Figure 2 to give Figure 4 we have the following boundary operator on Figure 4

$$\partial_1 = \begin{bmatrix} -\sqrt{\frac{m}{\Delta t}} & 0 & 0 & -\sqrt{-k\Delta t} & 0 & 0 & 0 \\ \sqrt{\frac{m}{\Delta t}} & -\sqrt{-k\Delta t} & -\sqrt{\frac{m}{\Delta t}} & 0 & 0 & 0 & 0 \\ 0 & 0 & \sqrt{\frac{m}{\Delta t}} & 0 & 0 & 0 & -\sqrt{-k\Delta t} \\ 0 & 0 & 0 & \sqrt{-k\Delta t} & -\sqrt{\frac{m}{\Delta t}} & 0 & 0 \\ 0 & \sqrt{-k\Delta t} & 0 & 0 & \sqrt{\frac{m}{\Delta t}} & -\sqrt{-k\Delta t} & 0 \\ 0 & 0 & 0 & 0 & 0 & \sqrt{-k\Delta t} & \sqrt{-k\Delta t} \end{bmatrix} \tag{6}$$

constructed analogously to Eq (2). One then finds *a la* Wise[(41)] that $\vec{\vec{K}} = \partial_1 \partial_1^T$. One can also read off the rows of $\vec{\vec{K}}$ by noting that row 1 says links of weight $\frac{m}{\Delta t}$ and $-k\Delta t$ are connecting nodes 1, 2 and 4, respectively. All other rows can be read off the same way.

---

[10] The column space is equal to the row space here, since $\vec{\vec{K}}$ is symmetric.

Either way, $\bar{\bar{K}}$ is understood to be constructed via graphical relations, so it might be called the "relations matrix."

Now that we have explained the AGC, our choice of gauge fixing is obvious. The discrete, graphical counterpart to Eq (4) is

$$Z = \int_{-\infty}^{\infty} \ldots \int_{-\infty}^{\infty} dQ_1 \ldots dQ_N \exp\left[ i\frac{1}{2}\vec{Q}\cdot\bar{\bar{K}}\cdot\vec{Q} + i\vec{J}\cdot\vec{Q} \right] \tag{7}$$

with solution

$$Z = \left( \frac{(2\pi i)^N}{\det(K)} \right)^{1/2} \exp\left[ -i\frac{1}{2}\vec{J}\cdot\bar{\bar{K}}^{-1}\cdot\vec{J} \right] \tag{8}$$

However, $\bar{\bar{K}}^{-1}$ does not exist because $\bar{\bar{K}}$ has a non-trivial null space. This is the graphical characterization of the effect of gauge invariance on the computation of $Z(J)$. Because we require that $\vec{J}$ reside in the row space of $\bar{\bar{K}}$, the graphical counterpart to Fadeev-Popov gauge fixing is clear, i.e., we simply restrict our path integral to the row space of $\bar{\bar{K}}$. Nothing of physical interest lies elsewhere, so this is a natural choice. In the eigenbasis of $\bar{\bar{K}}$ with our gauge fixing Eq (7) becomes

$$Z = \int_{-\infty}^{\infty} \ldots \int_{-\infty}^{\infty} d\tilde{Q}_2 \ldots d\tilde{Q}_N \exp\left[ \sum_{n=2}^{N} \left( i\frac{1}{2}\tilde{Q}_n^2 a_n + i\tilde{J}_n\tilde{Q}_n \right) \right] \tag{9}$$

where $\tilde{Q}_n$ are the coordinates associated with the eigenbasis of $\bar{\bar{K}}$ and $\tilde{Q}_1$ is associated with eigenvalue zero, $a_n$ is the eigenvalue of $\bar{\bar{K}}$ corresponding to $\tilde{Q}_n$, and $\tilde{J}_n$ are the components of $\vec{J}$ in the eigenbasis of $\bar{\bar{K}}$. Our gauge independent approach revises Eq. (8) to give

$$Z = \left( \frac{(2\pi i)^{N-1}}{\prod_{n=2}^{N} a_n} \right)^{1/2} \prod_{n=2}^{N} \exp\left[ -i\frac{\tilde{J}_n^2}{2a_n\hbar} \right] \tag{10}$$

*Thus, we find that the adynamically constrained co-construction of space, time and divergence-free sources entails gauge invariance and gauge fixing.* After quickly checking the general structure for unweighted scalar fields on the hypercube, we will

apply this idea to the Schrödinger, Klein-Gordon, Dirac, Maxwell, and Einstein-Hilbert actions.

*3.3 Unweighted Scalar Fields on the Hypercube.* We now provide $\bar{\bar{K}}1 = \partial_1\partial_1^T$, $\bar{\bar{K}}2 = \partial_2\partial_2^T$, $\bar{\bar{K}}3 = \partial_3\partial_3^T$, the eigenvalues for each $\bar{\bar{K}}$, and the structure of the row space for each $\bar{\bar{K}}$ on the hypercube (Figure 8) with unweighted links, plaquettes and cubes. These boundary operators satisfy $\partial_n\partial_{n+1} = 0$. For $\bar{\bar{K}}1 = \partial_1\partial_1^T$ the eigenvalues are {8,6,6,6,6,4,4,4,4,4,4,2,2,2,2,0} and the null space is span{$[111\ldots]^T$}, which we know from the fact that the rows of $\bar{\bar{K}}$ sum to zero. The AGC then means $\vec{J}$ sums to zero globally (all 16 nodes). For $\bar{\bar{K}}2 = \partial_2\partial_2^T$ the eigenvalues are {8,8,8,6,6,6,6,6,6,6,6,4,4,4,4,4,4,0,0,0,0,0,0,0,0,0,0,0,0,0,0,0}. The null space contains $[111\ldots]^T$, since the rows of $\bar{\bar{K}}$ sum to zero. The AGC then means $\vec{J}$ sums to zero globally (all 32 links). The null space contains vectors which correspond to $\vec{J}$ conserved on the links at each node (Figure 9). Thus, we can understand the 17 dimensions of the row space as resulting from degrees of freedom. If we specify $\vec{J}$ on all 12 links of the "inner" cube, all the time-like links connecting the "inner" cube to the "outer" cube are determined by local conservation. Then if you specify the 4 link values on one face of the "outer" cube, local conservation leaves only one free link to specify on the opposite face, 12 + 4 + 1 = 17. For $\bar{\bar{K}}3 = \partial_3\partial_3^T$ the eigenvalues are {8,8,8,6,6,6,6,0,0,0,0,0,0,0,0,0,0,0,0,0,0,0,0,0}. The null space contains $[111\ldots]^T$, since the rows of $\bar{\bar{K}}$ sum to zero. The AGC then means $\vec{J}$ sums to zero globally (all 24 plaquettes). There are vectors in the null space which correspond to $\vec{J}$ conserved on the plaquettes at each link (Figure 10). Thus, we can understand the 7 dimensions of the row space as follows. Start by specifying $\vec{J}$ on the six plaquettes of the "inner" cube. Then local conservation dictates the value of $\vec{J}$ on all plaquettes connecting the "inner" cube to the "outer" cube. That means we need only specify the value of $\vec{J}$ on one plaquette of the "outer" cube and local conservation will dictate the rest of its values on the "outer" cube. We next apply this approach to the free-particle Schrödinger action.

*3.4 Non-relativistic Scalar Field on Nodes and the Twin-Slit Experiment.* The non-relativistic limit of the Klein-Gordon (KG) equation gives the free-particle Schrödinger equation (SE) by factoring out the rest mass contribution to the energy *E*, assuming the Newtonian form for kinetic energy, and discarding the second-order time derivative[(42)]. To illustrate the first two steps, plug $\varphi = Ae^{i(px-Et)/\hbar}$ into the KG equation and obtain $\left(-E^2 + p^2c^2 + m^2c^4\right) = 0$, which tells us *E* is the total relativistic energy. Now plug $\psi = Ae^{i(px-Et)/\hbar}$ into the free-particle SE and obtain $\frac{p^2}{2m} = E$, which tells us *E* is only the Newtonian kinetic energy. Thus, we must factor out the rest energy of the particle, i.e., $\psi = e^{imc^2t/\hbar}\varphi$, assume the low-velocity limit of the relativistic kinetic energy, and discard the relevant term from our Lagrangian density (leading to the second-order time derivative) in going from $\varphi$ of the KG equation to $\psi$ of the free-particle SE. We will make these changes to *Z*(*J*) for the KG equation and obtain $\psi$(x,t), which we will then compare to $\psi$(x,t) from QM (with a source) to produce our probability amplitude.

For the KG equation we have

$$Z(J) = \int D\varphi \exp\left[i\int d^4x\left[\frac{1}{2}(\partial\varphi)^2 - \frac{1}{2}\overline{m}^2\varphi^2 + J\varphi\right]\right] \tag{11}$$

(overall factor of ħ in exponent is set to 1) which in (1+1)D is

$$Z(J) = \int D\varphi \exp\left[i\int dxdt\left[\frac{1}{2}\left(\frac{\partial\varphi}{\partial t}\right)^2 - \frac{c^2}{2}\left(\frac{\partial\varphi}{\partial x}\right)^2 - \frac{1}{2}\overline{m}^2\varphi^2 + J\varphi\right]\right] \tag{12}$$

($\overline{m} \equiv \frac{mc^2}{\hbar}$). Making the changes described above with $\psi = e^{i\overline{m}t}\sqrt{\overline{m}}\varphi$, Eq (12) gives the non-relativistic KG transition amplitude corresponding to the free-particle SE[(43)]

$$Z(J) = \int D\psi \exp\left[i\int dxdt\left[i\psi^*\left(\frac{\partial\psi}{\partial t}\right) - \frac{c^2}{2\overline{m}}\left(\frac{\partial\psi}{\partial x}\right)^2 + J\psi\right]\right] \tag{13}$$

Now integrate the second term by parts and obtain

$$Z(J) = \int D\psi \exp\left[i\int dxdt\left[i\psi^*\left(\frac{\partial\psi}{\partial t}\right) + \frac{\hbar}{2m}\psi^*\frac{\partial^2\psi}{\partial x^2} + J\psi\right]\right] \tag{14}$$

This gives

$$Z(J) = \int D\psi \exp\left[i\int dxdt\left[\frac{1}{2}\psi^* K\psi + J\psi\right]\right] \tag{15}$$

where

$$K = \left(2i\frac{\partial}{\partial t} + \frac{\hbar}{m}\frac{\partial^2}{\partial x^2}\right) \tag{16}$$

The solution to this is

$$Z(J) = Z(0)\exp\left(-\frac{i}{2}\iint dxdx'J(x)D(x-x')J(x')\right) \equiv Z(0)\exp(iW(J)) \tag{17}$$

where $x$ and $x'$ are each shorthand for both a spatial dimension and a temporal dimension,

$$W(J) = -\frac{1}{2}\iint dxdx'J(x)D(x-x')J(x') \tag{18}$$

and

$$\left(2i\frac{\partial}{\partial t} + \frac{\hbar}{m}\frac{\partial^2}{\partial x^2}\right)D(x-x') = \delta(x-x') \tag{19}$$

that is, $D(x - x')$ is the Green's function, aka the QFT propagator. A solution is

$$D(x-x') = \frac{-1}{(2\pi)^2}\iint \frac{e^{ik(x-x')}e^{i\omega(t-t')}d\omega dk}{2\omega + \hbar k^2/m} \tag{20}$$

(where $x$ and $x'$ are just spatial on the RHS). Notice from Eq (17) that $D(x - x')$ *is worthless in the absence of a source*. This is important in an RBW approach, since Nature is COs defined relationally/contextually via "quantum interactions" there is no truly "sourceless" physics.

With a source, one uses $D(x - x')$ to solve the differential equation

$$\hat{\Theta}_x\phi(x) = S(x) \tag{21}$$

by writing

$$\phi(x) = \int dx' D(x-x')S(x') \tag{22}$$

where

$$\hat{\Theta}_x D(x-x') = \delta(x-x') \tag{23}$$

since

$$\hat{\Theta}_x \phi(x) = \int dx' \hat{\Theta}_x D(x-x')S(x') = S(x) \tag{24}$$

That $D(x - x')$ is worthless without sources is significant because the QM free-particle propagator[44] with $\psi(x,0) = \delta(x)$ yields

$$\psi_o(x,t) = \sqrt{\frac{m}{2\pi\hbar i t}} \exp\left[\frac{imx^2}{2\hbar t}\right] \tag{25}$$

and this gives

$$\left(2i\frac{\partial}{\partial t} + \frac{\hbar}{m}\frac{\partial^2}{\partial x^2}\right)\psi_o(x,t) = 0 \tag{26}$$

Thus, $\psi_o(x,t)$ obtained from the QM free-particle propagator is a solution of the SE without a source, i.e, $S(x) = 0$ in Eq (21) and $J = 0$ in Eq (17). So, QM's $\psi_o(x,t)$ in Eq (25) is not physically relevant in RBW per QFT's $Z(J)$. That is, since $Z(J) = Z(0)e^{iW(J)}$ , the sourceless solutions $\psi_o(x,t)$ would appear in the exponent of $Z(0)$ which Zee describes as[45], "often of no interest to us." In order to obtain a physically relevant "free-particle amplitude" related to non-relativistic QFT, the SE must have a "source." Essentially, in our QFT approach, we want a particle of mass $m$ created at the Source and annihilated at the sink (detector) – with no worldline connecting them – and this happens at sources $J$. We can write the SE[46]

$$\left(2i\frac{\partial}{\partial t} + \frac{\hbar}{m}\frac{\partial^2}{\partial x^2}\right)\psi(x) = 2V(x)\psi(x) \tag{27}$$

so that Eq (22) gives us

$$\psi(x) = \int dx' D(x-x')2V(x')\psi(x') \tag{28}$$

With $2V(x')\psi(x') = S(x') = \delta(x'-x_i)$ we have

$$\psi(x) = D(x-x_i) \tag{29}$$

We could still add solutions $\psi_o(x)$ of the sourceless equation, but again they are associated with $Z(0)$ and therefore of "no interest to us."

To find the QFT counterpart to Eq (29), we use Eq (18) with point sources $J(x')$ at $x_i$ (Source) and $J(x)$ at $x_f$ (sink/detector) to obtain

$$W(J) = -\frac{1}{2}\iint dxdx'\delta(x - x_f)D(x - x')\delta(x' - x_i) = -\frac{1}{2}D(x_f - x_i) \qquad (30)$$

So, with $D(x - x')$ given by Eq (17) we have our QFT derivation of the "free-particle" QM probability amplitude, i.e., $\psi(x) = -2W(J)$, which is $\vec{J}\cdot\bar{\bar{K}}^{-1}\cdot\vec{J}$ on the graph of MLGT. That we must always supply $J(x)$, and that $J(x)$ is always coupled to $J(x')$ via $D(x - x')$ in $Z(J)$, is consistent with the relational ontology of RBW. Now we formulate the discrete MLGT counterpart to this result.

Since $\psi^*$ appears undifferentiated in Eq (13), we do not have a fully relational form. We imagine that this is because $\psi$ needs to be underwritten by a "coordinate" field that reveals the underlying relational form of the action. For example, if one writes the spring potential of section 3.2 in terms of the displacement $x$ from equilibrium, one obtains the term $\frac{1}{2}kx^2$ in the action, but this obscures the relational structure revealed using coordinates $q$, i.e., $\frac{1}{2}k(q_1 - q_2)^2$. So, we replace $\psi^*$ with a relational structure $\psi^* \rightarrow (\psi_2^* - \psi_1^*)$ in the following discretizations (with extrapolations):

$$i\psi^*\left(\frac{\partial\psi}{\partial t}\right) \rightarrow i(\psi_2^* - \psi_1^*)\left(\frac{\psi_2 - \psi_1}{\Delta t}\right)$$

$$-\frac{c^2}{2\overline{m}}\left(\frac{\partial\psi}{\partial x}\right)^2 \rightarrow -\frac{\hbar}{2m}\left(\frac{\psi_3^* - \psi_1^*}{\Delta x}\right)\left(\frac{\psi_3 - \psi_1}{\Delta x}\right)$$

where $\psi_2$ is at node $\psi_1 + \Delta t$, $\psi_3$ is at node $\psi_1 + \Delta x$, and $\psi_4$ is at node $\psi_1 + \Delta x + \Delta t$ (Figure 5). We obtain for $\bar{\bar{K}}$ in $\frac{1}{2}\vec{\psi}^*\cdot\bar{\bar{K}}\cdot\vec{\psi}$ :

$$\overset{=}{K}=\begin{bmatrix} \left(\frac{2i}{\Delta t}-\frac{\hbar}{m\Delta x^2}\right) & -\frac{2i}{\Delta t} & \frac{\hbar}{m\Delta x^2} & 0 \\ -\frac{2i}{\Delta t} & \left(\frac{2i}{\Delta t}-\frac{\hbar}{m\Delta x^2}\right) & 0 & \frac{\hbar}{m\Delta x^2} \\ \frac{\hbar}{m\Delta x^2} & 0 & \left(\frac{2i}{\Delta t}-\frac{\hbar}{m\Delta x^2}\right) & -\frac{2i}{\Delta t} \\ 0 & \frac{\hbar}{m\Delta x^2} & -\frac{2i}{\Delta t} & \left(\frac{2i}{\Delta t}-\frac{\hbar}{m\Delta x^2}\right) \end{bmatrix} \tag{31}$$

ignoring the volume element $\Delta x\Delta t$. The eigenvalues are $\{0, \frac{4i}{t}, -\frac{2\hbar}{mx^2}, \frac{4i}{t}-\frac{2\hbar}{mx^2}\}$, where we have dropped the $\Delta$ for simplicity, and the corresponding eigenvectors are {(1,1,1,1), (-1,1,-1,1), (-1,-1,1,1), (1,-1,-1,1)}, i.e., a Hadamard structure we will see repeated in both the KG and Dirac equations. These eigenvectors correspond to the following four modes, respectively:

**Mode 1**

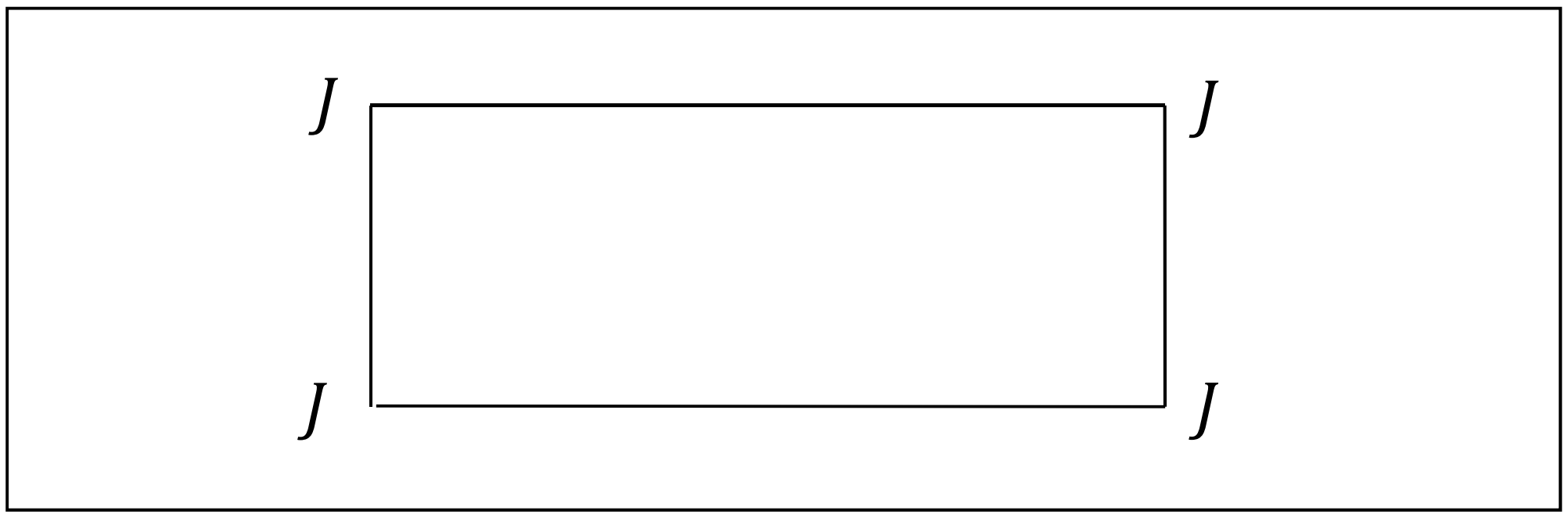

There is no spatial or temporal variation in $\vec{J}$, so $\vec{J}$ is not divergence-free and therefore does not reside in the row space of $\overset{=}{K}$. This source does not satisfy the AGC.

**Mode 2**

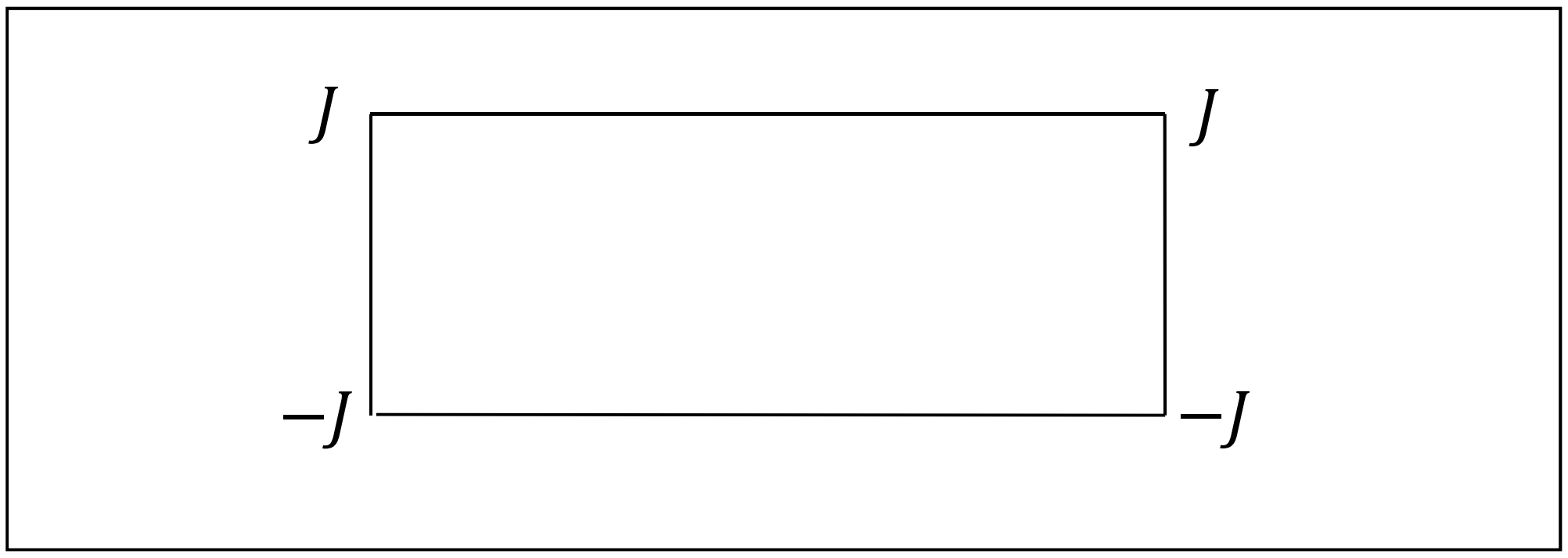

There is only temporal variation in $\vec{J}$. While $\vec{J}$ resides in the row space of $\vec{\vec{K}}$ and is therefore divergence-free in the mathematical sense, it is not conserved within the element. Therefore, this source does not satisfy the AGC.

**Mode 3**

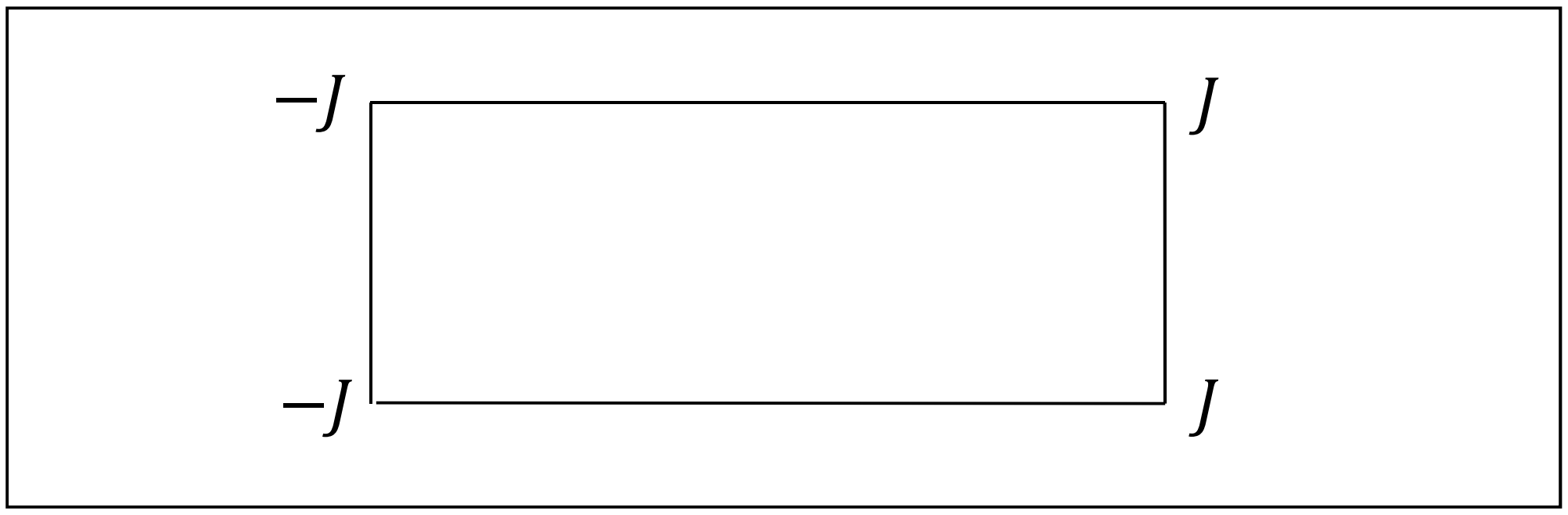

There is only spatial variation in $\vec{J}$. While $\vec{J}$ resides in the row space of $\vec{\vec{K}}$ and is conserved within the element, it does not represent an interaction. Therefore, this source does not satisfy the AGC.

**Mode 4**

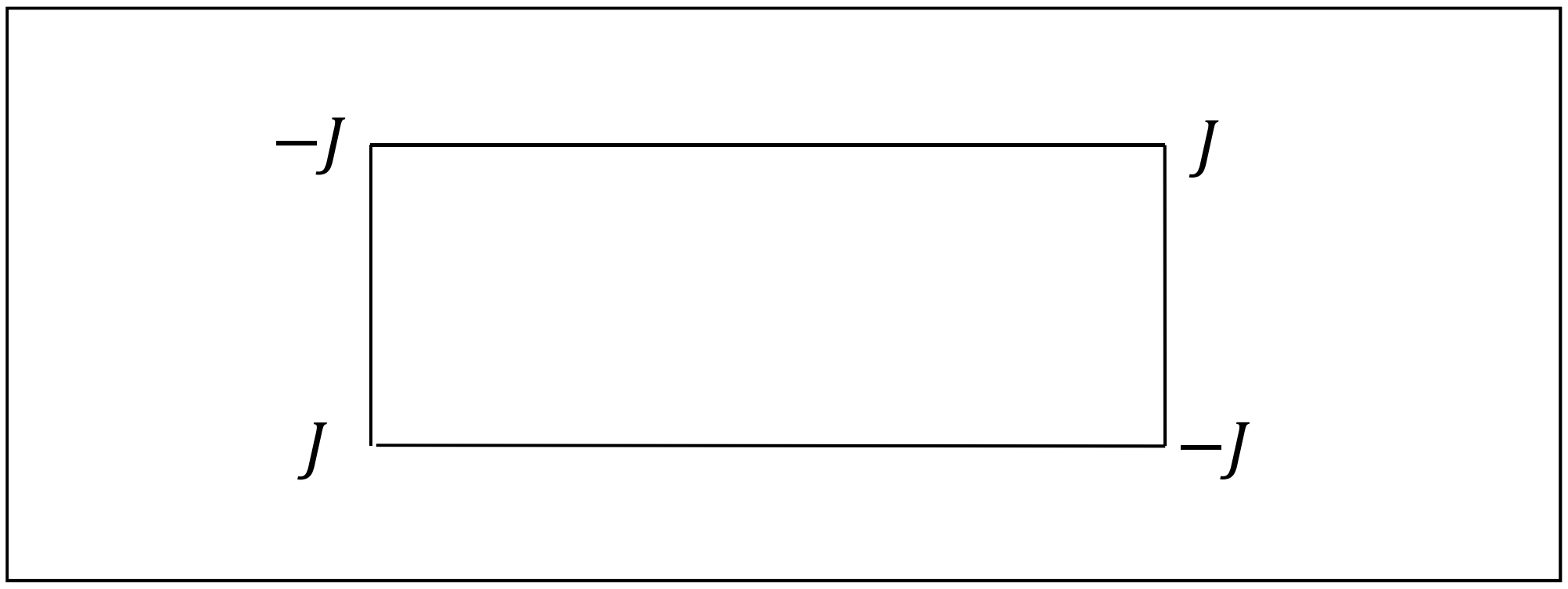

There is both spatial and temporal variation in $\vec{J}$, which resides in the row space of $\vec{\vec{K}}$, is conserved in the element, and represents an interaction. This source satisfies the AGC.

In the eigenspace of $\vec{\vec{K}}$, the source associated with mode 4 is $\vec{J} = (0, 0, 0, J_o)$, where $J_o$ is complex in general, so it is easily seen that (with our gauge fixing)

$$\vec{J} \cdot \vec{\vec{K}}^{-1} \cdot \vec{J} = \frac{J_o^2}{\left(\frac{4i}{t} - \frac{2\hbar}{mx^2}\right)} \tag{32}$$

Eqs (29) and (30) tell us that Eq (32) is the MLGT counterpart to Eq (20), i.e.,

$$\frac{J_o^2}{\left(\frac{4i}{t} - \frac{2\hbar}{mx^2}\right)} = \frac{-1}{(2\pi)^2} \iint \frac{e^{ikx} e^{i\omega t} d\omega dk}{2\omega + \hbar k^2 / m} \tag{33}$$

where $t$ and $x$ represent the temporal and spatial extent of the element, respectively, and $J_o^2 = J_o \cdot J_o$ (not $J_o^* \cdot J_o$). The LHS of Eq (33) simply explains the graphical origin of the RHS which gives

$$-\frac{1}{4}\sqrt{\frac{m}{\pi\,\hbar\,t}}\left[i\mathrm{C}\left(\sqrt{\frac{m\,x^2}{\pi\,\hbar\,t}}\right) + \mathrm{S}\left(\sqrt{\frac{m\,x^2}{\pi\,\hbar\,t}}\right)\right]\exp\left(\frac{i\,m\,x^2}{2\,\hbar\,t}\right)$$

where $C(z) = \int_0^z \cos\left(\frac{\pi}{2}u^2\right) du$ and $S(z) = \int_0^z \sin\left(\frac{\pi}{2}u^2\right) du$ are Fresnel integrals. Let us denote this A($x$, $t$, $m$). Now to construct the amplitude $A_{total}$ for a spacetimesource element for an outcome in the twin-slit experiment, we have (Figure 6):

$$A_{total} = A(x_1, t_1, m)A(x_3, t_3, m) + A(x_2, t_2, m)A(x_4, t_4, m)$$

where $x_1$ and $t_1$ are the distance and time from Source to Slit 1, $x_2$ and $t_2$ are the distance and time from Source to Slit 2, $x_3$ and $t_3$ are the distance and time from Slit 1 to the Detection Event, and $x_4$ and $t_4$ are the distance and time from Slit 2 to the Detection Event. For an electron traveling at 1.00 m/s through the device (dynamic language) we obtain the following plot. [Note: The amplitudes of Eqs (25) and (33) were computed for the properties of space, time and mass. In order to model the data for their twin-slit experiment with electrons, Bach *et al.* [47] had to modify the "free space" amplitude to include other properties. Modifications included an electromagnetic potential at the double slits, an image charge potential at the collimation slit, and incoherent sources associated with the electron gun. Therefore, differences in the plots below are not expected to be experimentally observable for electrons (the angles shown below exceed

±π, for example). The point of this exercise is only to illustrate the manner by which RBW underwrites QM via the AGC.]

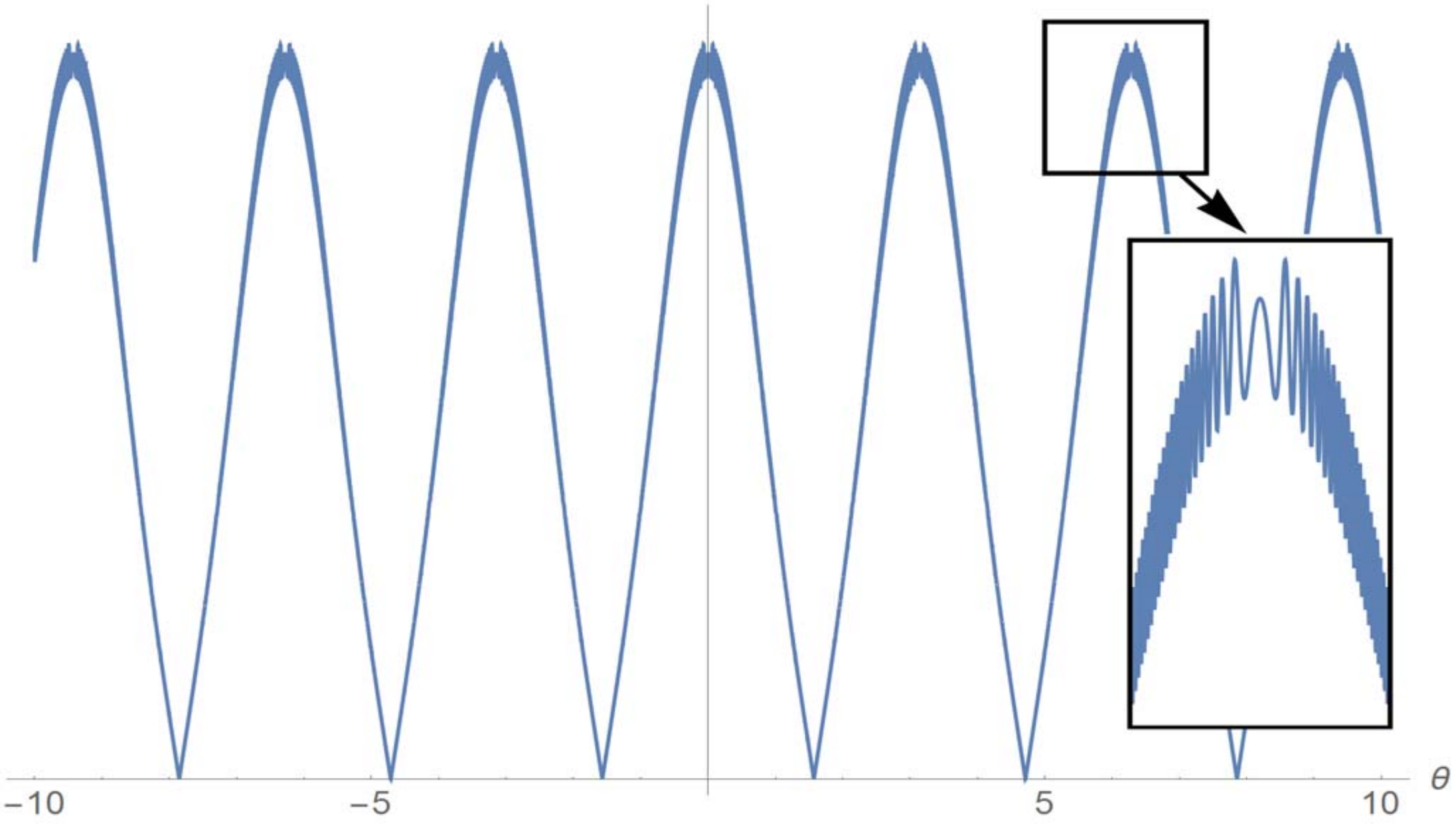

Intensity versus angular displacement in radians for electrons with λ = 728 μm, slit separation of 1.00 mm, screen-to-detector distance of 50.0 cm, and Source-to-slits distance of 50.0 cm. This is the RBW result. There is an oscillatory substructure that is suppressed by the horizontal scale (see inset).

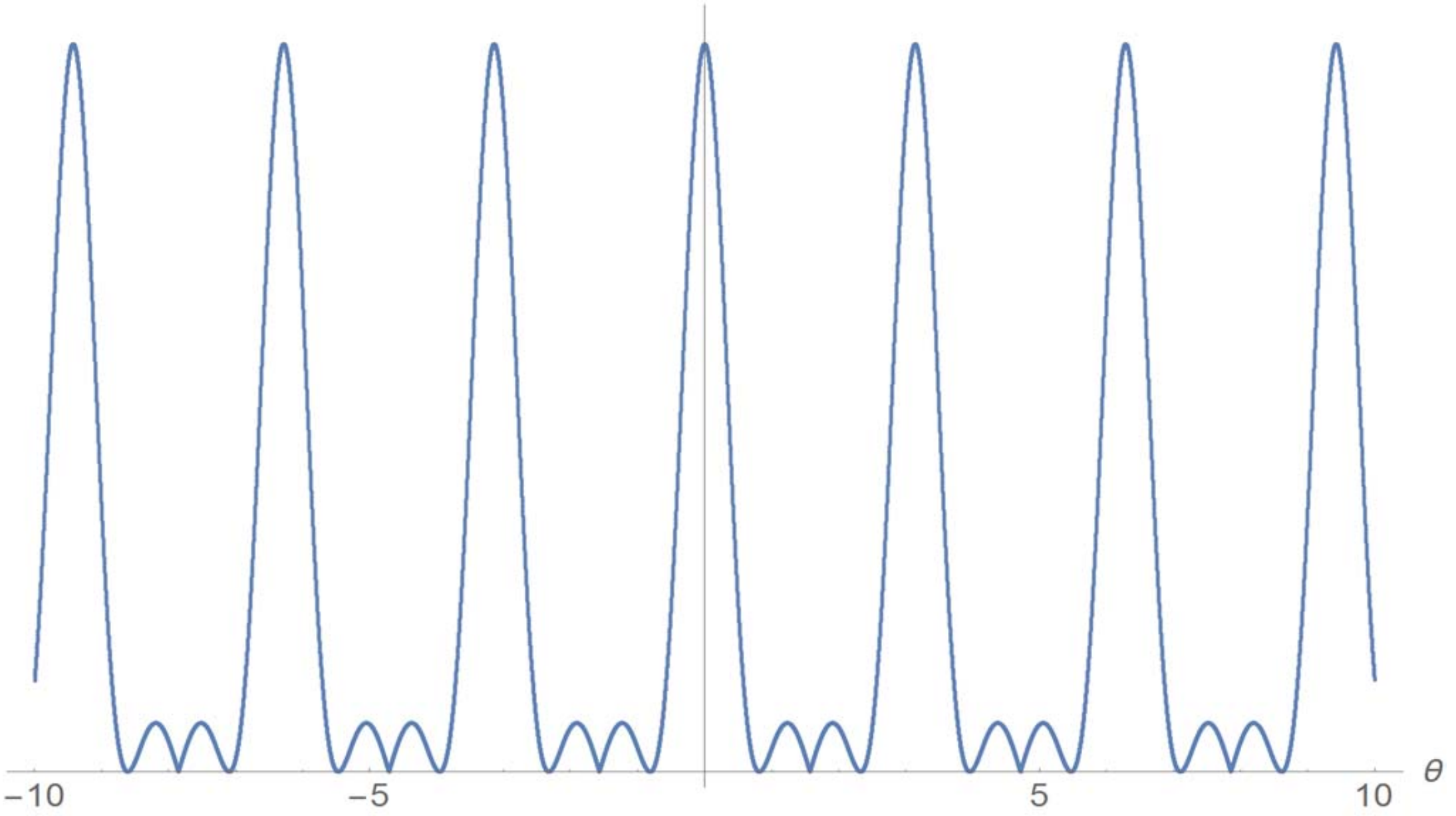

Intensity versus angular displacement in radians for electrons with λ = 728 μm, slit separation of 1.00 mm, screen-to-detector distance of 50.0 cm, and Source-to-slits distance of 50.0 cm. This is the free-particle SE result without a source given by Eq (25). Large maxima align with those in the RBW result above at this very low energy.

In the standard account of twin-slit interference, quantum waves emanate from a pair of coherent Sources and strike all along the detector surface (Figure 7). Exactly how these waves produce a single point on the detector surface is then left to interpretation, e.g., they actually "guide" quantum particles per the deBroglie-Bohm interpretation. The standard account is a very dynamical view of twin-slit interference. By contrast, the 4-component spacetimesource element of Figure 6 is obtained in spatiotemporally global fashion, as we described in section 1, i.e., the entire 4-component spacetimesource element is a fundamental, indivisible, ontological subset of the interacting Source, screen, and detector (again, the properties of classical objects are relational). There is no worldline connecting the emission event at the Source and the Detection Event. You can think of this spacetimesource element as just one of many responsible for the Source, screen, and detector, to include their spatiotemporal properties. This understanding of quantum exchanges, interference, and spacetimesource elements, combined with our view of particle physics (section 4.2), will explain why the Standard Model Lagrangian density is so complicated (Figure 12) and why it's not fundamental. We next study the Klein-Gordon action and find it shares the Hadamard structure of the Schrödinger result.

*3.5 Scalar Field on Nodes.* We now consider Eq (12). The 4-node graph of Figure 5 with $\psi$ replaced by $\varphi$ depicts our spacetimesource element for this case. Again, we have an undifferentiated field in the action so we will have to replace it with a coordinate form. To mirror the SE discretization we choose $(\varphi_4^* - \varphi_3^*)(\varphi_4 - \varphi_3) + (\varphi_2^* - \varphi_1^*)(\varphi_2 - \varphi_1)$ for the discretization of $\varphi^2$ in our spacetimesource element. This gives (having factored out ½ and ignored ΔxΔt)

$$\vec{\vec{K}} = \begin{bmatrix} \left(\frac{1}{\Delta t^2} - \frac{c^2}{\Delta x^2} - \overline{m}^2\right) & \frac{1}{\Delta t^2} - \overline{m}^2 & \frac{c^2}{\Delta x^2} & 0 \\ \frac{1}{\Delta t^2} - \overline{m}^2 & \left(\frac{1}{\Delta t^2} - \frac{c^2}{\Delta x^2} - \overline{m}^2\right) & 0 & \frac{c^2}{\Delta x^2} \\ \frac{c^2}{\Delta x^2} & 0 & \left(\frac{1}{\Delta t^2} - \frac{c^2}{\Delta x^2} - \overline{m}^2\right) & \frac{1}{\Delta t^2} - \overline{m}^2 \\ 0 & \frac{c^2}{\Delta x^2} & \frac{1}{\Delta t^2} - \overline{m}^2 & \left(\frac{1}{\Delta t^2} - \frac{c^2}{\Delta x^2} - \overline{m}^2\right) \end{bmatrix} \quad (34)$$

The eigenvalues are $\{0,\ \frac{2}{t^2}-\frac{2m^2c^4}{\hbar^2},-\frac{2c^2}{x^2},-\frac{2c^2}{x^2}+\frac{2}{t^2}-\frac{2m^2c^4}{\hbar^2}\}$, where we have dropped the Δ for simplicity, and the corresponding eigenvectors are Hadamard just as with the SE {(1,1,1,1), (-1,1,-1,1), (-1,-1,1,1), (1,-1,-1,1)}. Again, mode 4 is the $\vec{J}$ that satisfies our AGC so our amplitude is

$$\vec{J}\cdot\bar{\bar{K}}^{-1}\cdot\vec{J}=\frac{J_o^2}{\left(-\frac{2c^2}{x^2}+\frac{2}{t^2}-\frac{2m^2c^4}{\hbar^2}\right)} \tag{35}$$

We next study the Dirac action and find it also shares the Hadamard structure.

*3.6 Vector Field on Nodes*. We apply this approach to vector fields on nodes and note that the KG operator for scalar fields is the square of the Dirac operator for vector fields, i.e., $\left(-i\gamma^\mu\partial_\mu-m\right)\left(i\gamma^\mu\partial_\mu-m\right)=\left(\partial^2+m^2\right)$. In order to construct $\bar{\bar{K}}$ for the Dirac operator on the hypercube of Figure 8 we have the following link weights on *t*, *x*, *y*, and *z* links respectively:

$$T=\begin{bmatrix}\frac{1}{t}-m & 0 & 0 & 0\\ 0 & \frac{1}{t} & 0 & 0\\ 0 & 0 & \frac{1}{t} & 0\\ 0 & 0 & 0 & \frac{1}{t}\end{bmatrix}\qquad X=\begin{bmatrix}0 & 0 & 0 & \frac{1}{x}\\ 0 & -m & \frac{1}{x} & 0\\ 0 & \frac{1}{x} & 0 & 0\\ \frac{1}{x} & 0 & 0 & 0\end{bmatrix} \tag{36}$$

$$Y=\begin{bmatrix}0 & 0 & 0 & -\frac{i}{y}\\ 0 & 0 & \frac{i}{y} & 0\\ 0 & -\frac{i}{y} & m & 0\\ \frac{i}{y} & 0 & 0 & 0\end{bmatrix}\qquad Z=\begin{bmatrix}0 & 0 & \frac{1}{z} & 0\\ 0 & 0 & 0 & -\frac{1}{z}\\ \frac{1}{z} & 0 & 0 & 0\\ 0 & -\frac{1}{z} & 0 & m\end{bmatrix}$$

Then the 64 x 64 matrix $\bar{\bar{K}}$ is simply given by:

$$\bar{\bar{K}} = \begin{bmatrix} (-T-X-Y-Z) & Z & Y & 0 & X & \dots \\ & \vdots & & & & \end{bmatrix} \tag{37}$$

This has the same form as $\bar{\bar{K}}$ for the Schrödinger and KG actions. That is, reading across the rows for each node one simply has a collection of the link weights relating the nodes which are connected. Thus, as claimed in section 2, we can understand how $\bar{\bar{K}}$ instantiates graphical relationalism and divergence-free $\vec{J}$ per the AGC.

To study the eigenstructure, we point out that $\bar{\bar{K}}$ is in nested form. $\bar{\bar{K}}_{block} = \begin{bmatrix} A & TI \\ TI & A \end{bmatrix}$ where $TI$ is the 8x8 identity matrix $I$ times $T$ and $A$ is the 8x8 matrix $A = \begin{bmatrix} B & XI \\ XI & B \end{bmatrix}$.

Continuing the nesting we have $B = \begin{bmatrix} C & YI \\ YI & C \end{bmatrix}$ where $C = \begin{bmatrix} D & ZI \\ ZI & D \end{bmatrix}$ and $D = [-T-X-Y-Z]$. The eigenvalue problem for $\bar{\bar{K}}$ then takes a nested form in terms of Hadamard matrices $H_1$, $H_2$, $H_4$, $H_8$, and $H_{16}$ as follows. $DH_1 = H_1[-T-X-Y-Z]$ where $H_1 = [1]$. $C\begin{bmatrix} 1 & 1 \\ 1 & -1 \end{bmatrix} = \begin{bmatrix} 1 & 1 \\ 1 & -1 \end{bmatrix}\begin{bmatrix} -T-X-Y & 0 \\ 0 & -T-X-Y-2Z \end{bmatrix}$ where $H_2 = \begin{bmatrix} 1 & 1 \\ 1 & -1 \end{bmatrix}$.

$BH_4 = H_4 diag[-T-X, -T-X-2Z, -T-X-2Y, -T-X-2Y-2Z]$ where

$$H_4 = \begin{bmatrix} 1 & 1 & 1 & 1 \\ 1 & -1 & 1 & -1 \\ 1 & 1 & -1 & -1 \\ 1 & -1 & -1 & 1 \end{bmatrix}.$$

$$AH_8 = H_8 diag[-T, -T-2Z, -T-2Y-2Z, -T-2X, -T-2X-2Z, -T-2X-2Y, -T-2X-2Y-2Z]$$

where $H_8 = \begin{bmatrix} H_4 & H_4 \\ H_4 & -H_4 \end{bmatrix} = H_2 \otimes H_4$. Thus, $\bar{\bar{K}}_{block} H_{16} = H_{16} diag[vector]$ where

$H_{16} = H_2 \otimes H_8$ and

$$vector = -2\begin{bmatrix} 0 \\ Z \\ Y \\ Y+Z \\ X \\ X+Z \\ X+Y \\ X+Y+Z \\ T \\ T+Z \\ T+Y \\ T+Y+Z \\ T+X \\ T+X+Z, \\ T+X+Y \\ T+X+Y+Z \end{bmatrix}$$

Finally, the eigenvalue problems for each of the 4x4 matrices in *vector* are solved and the eigenvectors are located in a 64x64 matrix built from $H_{16}$. So, for example, the first column of $H_{16}$ is $[111\ldots]^T$ and the four-dimensional null space is spanned by [1,0,0,0], [0,1,0,0], [0,0,1,0] and [0,0,0,1], so the first four columns of the eigenbasis matrix for $\vec{\vec{K}}$ are (column entries top to bottom read left to right here):

```
{{0, 0, 0, 1, 0, 0, 0, 1, 0, 0, 0, 1, 0, 0, 0, 1, 0, 0, 0, 1, 0, 0, 0, 1, 0, 0, 0, 1, 0, 0, 0, 1, 0, 0, 0, 1, 0, 0, 0, 1, 0, 0,
  0, 1, 0, 0, 0, 1, 0, 0, 0, 1, 0, 0, 0, 1, 0, 0, 0, 1, 0, 0, 0, 1}, {0, 0, 1, 0, 0, 0, 1, 0, 0, 0, 1, 0, 0, 0, 1, 0, 0, 0, 1, 0, 0,
  0, 1, 0, 0, 0, 1, 0, 0, 0, 1, 0, 0, 0, 1, 0, 0, 0, 1, 0, 0, 0, 1, 0, 0, 0, 1, 0, 0, 0, 1, 0, 0, 0, 1, 0, 0, 0, 1, 0, 0, 0, 1, 0},
 {0, 1, 0, 0, 0, 1, 0, 0, 0, 1, 0, 0, 0, 1, 0, 0, 0, 1, 0, 0, 0, 1, 0, 0, 0, 1, 0, 0, 0, 1, 0, 0, 0, 1, 0, 0, 0, 1, 0, 0, 0, 1, 0,
  0, 0, 1, 0, 0, 0, 1, 0, 0, 0, 1, 0, 0, 0, 1, 0, 0, 0, 1, 0, 0}, {1, 0, 0, 0, 1, 0, 0, 0, 1, 0, 0, 0, 1, 0, 0, 0, 1, 0, 0, 0, 1,
  0, 0, 0, 1, 0, 0, 0, 1, 0, 0, 0, 1, 0, 0, 0, 1, 0, 0, 0, 1, 0, 0, 0, 1, 0, 0, 0, 1, 0, 0, 0, 1, 0, 0, 0, 1, 0, 0, 0, 1, 0, 0, 0}}
```

$\vec{J}$ being orthogonal to each of these vectors simply means that the global sum over each spacetime component of $\vec{J}$ at each node gives zero, as required for vector addition over all 16 nodes. We next study the Maxwell action.

*3.7 Scalar Field on Links.* We now apply this approach to gauge fields for the exchange of energy via photons. In order to model the construct of action for the exchange of energy via photons, we use the Maxwell Lagrangian density $L$ for free electromagnetic radiation

$$L = -\frac{1}{4\mu_o} F^{\alpha\beta} F_{\alpha\beta} \tag{38}$$

with the field strength tensor given by

$$F_{\alpha\beta} = \partial_\alpha A_\beta - \partial_\beta A_\alpha = \left[ \frac{\left(A_\beta(n+\hat{\alpha}) - A_\beta(n)\right)}{\ell_\alpha} - \frac{\left(A_\alpha(n+\hat{\beta}) - A_\alpha(n)\right)}{\ell_\beta} \right] \tag{39}$$

on the graph[48] where $n$ is the node number, $\ell_i$ the lattice spacing in the i$^{th}$ direction, and $\hat{\alpha}$ and $\hat{\beta}$ are displacements to adjoining nodes in those directions. Applying this to the (1+1)D case $\bar{\bar{K}}$ has eigenvalues 0, 0, 0, $2\left(\frac{1}{x^2} + \frac{1}{t^2}\right)$. The dimensionality of the row space represents the degrees of freedom available with local conservation of $\vec{J}$, as explained in section 3.3. That is, specifying $\vec{J}$ on just one link dictates the other three values per conservation of $\vec{J}$ on the links at each node.

On the cube, $\bar{\bar{K}}$ has eigenvalues

$$\{0,0,0,0,0,0,0,-\frac{2(t^2+x^2)}{t^2x^2},-\frac{2(t^2+y^2)}{t^2y^2},-\frac{2(x^2+y^2)}{x^2y^2},-\frac{2(t^2x^2+t^2y^2+x^2y^2)}{t^2x^2y^2},-\frac{2(t^2x^2+t^2y^2+x^2y^2)}{t^2x^2y^2}\}$$

of a combinatorial nature analogous to (1+1)D. Again, the dimensionality of the row space (five) represents the degrees of freedom available with local conservation of $\vec{J}$. That is, specifying $\vec{J}$ on the four links of one face (front, say) gives $\vec{J}$ on the links connecting the front face to the back face by local conservation. Then specifying $\vec{J}$ on just one link of the back face specifies the remaining links by local conservation.

$\bar{\bar{K}}$ for the hypercube has eigenvalues

$$\left\{\{0,0,0,0,0,0,0,0,0,0,0,0,0,0,0\},\left\{-\frac{2}{t^2}-\frac{2}{x^2}\right\},\left\{-\frac{2}{t^2}-\frac{2}{y^2}\right\},\left\{\frac{2}{x^2}+\frac{2}{y^2}\right\},\left\{\frac{2(t^2-x^2)}{t^2x^2}+\frac{2}{y^2}\right\},\left\{-\frac{2(t^2+x^2)}{t^2x^2}-\frac{2}{y^2}\right\},\left\{-\frac{2}{t^2}-\frac{2}{z^2}\right\},\left\{\frac{2}{x^2}+\frac{2}{z^2}\right\},\left\{\frac{2(t^2-x^2)}{t^2x^2}+\frac{2}{z^2}\right\},\left\{-\frac{2(t^2+x^2)}{t^2x^2}-\frac{2}{z^2}\right\},\left\{\frac{2}{y^2}+\frac{2}{z^2}\right\},\left\{\frac{2(t^2-y^2)}{t^2y^2}+\frac{2}{z^2}\right\},\left\{-\frac{2(t^2+y^2)}{t^2y^2}-\frac{2}{z^2}\right\},\left\{\frac{2(x^2+y^2)}{x^2y^2}+\frac{2}{z^2},\frac{2(x^2+y^2)}{x^2y^2}+\frac{2}{z^2}\right\},\left\{\frac{2(t^2x^2+t^2y^2-x^2y^2)}{t^2x^2y^2}+\frac{2}{z^2},\frac{2(t^2x^2+t^2y^2-x^2y^2)}{t^2x^2y^2}+\frac{2}{z^2}\right\},\left\{-\frac{2(t^2x^2+t^2y^2+x^2y^2)}{t^2x^2y^2}-\frac{2}{z^2}\right\}\right\}$$

of a combinatorial nature akin to the lower-dimensional versions. Again, the dimensionality of the row space (17) represents the degrees of freedom available with local conservation of $\bar{J}$, as explained in section 3.3 for links of the hypercube. We next study the Einstein-Hilbert action.

*3.8 Scalar Field on Plaquettes.* This is linearized GR, i.e., the harmonic terms only. We have for the Einstein-Hilbert Lagrangian density[49]

$$L = -\partial_\lambda h_{\alpha\beta}\partial^\lambda h^{\alpha\beta} + 2\partial_\lambda h_{\alpha\beta}\partial^\beta h^{\alpha\lambda} \qquad (40)$$

omitting a trace term not relevant to our application[11]. To discretize this on the hypercube (Figure 8) we first label our scalar field on each plaquette according to its span. For example, the front face of the "inner" cube is spanned by *x* and *z*, so it's labeled $h_{13}$. Of course, there are three other such plaquettes, one displaced from the front towards the back (in *y*) of the "inner" cube, one displaced in *t* to the front of the "outer" cube, and one displaced in *t* and *y* to the back of the "outer" cube. There are six fields ($h_{01}$, $h_{02}$, $h_{03}$, $h_{12}$, $h_{13}$, $h_{23}$) which generate such a quadruple, accounting for all 24 plaquettes of the hypercube. Likewise, for the cube we have ($h_{01}$, $h_{02}$, $h_{12}$) and their pairing partners giving us the six plaquettes.

[11] The missing trace term is gauge equivalent to $2\partial_\alpha h^\alpha_\mu$ which would be used for multiple, connected graphical elements.

We see that the first term of $L$ is just the sum of the squares of the gradients formed in each set of $h_{\alpha\beta}$ values, e.g.,

$$\left(\frac{h_{13}(back-in)}{y}-\frac{h_{13}(front-in)}{y}\right)^2+\left(\frac{h_{13}(back-out)}{y}-\frac{h_{13}(front-out)}{y}\right)^2+$$

$$\left(\frac{h_{13}(back-out)}{ct}-\frac{h_{13}(back-in)}{ct}\right)^2+\left(\frac{h_{13}(front-out)}{ct}-\frac{h_{13}(front-in)}{ct}\right)^2$$

for $h_{13}$ where "in" stands for "inner" cube and "out" stands for "outer" cube. The second term of $L$ is formed by mixing gradients, just as with the photon field in section 3.7. For example, we would have terms like $(\partial_0 h_{12})(\partial_2 h_{10})$ which on the lattice would have forms such as

$$\left(\frac{h_{12}(bottom-out)}{t}-\frac{h_{12}(bottom-in)}{t}\right)\left(\frac{h_{10}(back-in)}{y}-\frac{h_{10}(front-in)}{y}\right)$$

Using these conventions on the cube (again, ignoring overall scaling factors and letting $c = 1$), $\bar{\bar{K}}$ has eigenvalues

$$\{0,0,0,2(\frac{1}{x^2}+\frac{1}{y^2}),\frac{xy-\sqrt{x^2y^2+4t^2(x^2+y^2)}}{t^2xy},\frac{xy+\sqrt{x^2y^2+4t^2(x^2+y^2)}}{t^2xy}\}$$

and a basis for the null space is

$$\{\{0,0,0,0,1,1\},\{0,0,1,1,0,0\},\{1,1,0,0,0,0\}\}$$

which represents conservation of $\vec{J}$ among each pair of plaquettes associated with ($h_{01}$, $h_{02}$, $h_{12}$). [Of course, the rows of $\bar{\bar{K}}$ sum to zero so, as always, $[111\ldots]^T$ is a null eigenvector meaning we have global conservation of $\vec{J}$.] One null eigenbasis for $\bar{\bar{K}}$ of the hypercube is

$$\{\{0,0,0,0,0,0,0,0,0,0,0,0,0,0,0,0,0,0,0,0,1,1,1,1\},\{0,0,0,0,0,0,0,0,0,0,0,0,0,0,0,0,1,1,1,1,0,0,0,0\},$$
$$\{0,0,0,0,0,0,0,0,0,0,0,0,1,1,1,1,0,0,0,0,0,0,0,0\},\{0,0,0,0,0,0,0,0,1,1,1,1,0,0,0,0,0,0,0,0,0,0,0,0\},$$
$$\{0,0,0,0,1,1,1,1,0,0,0,0,0,0,0,0,0,0,0,0,0,0,0,0\},\{1,1,1,1,0,0,0,0,0,0,0,0,0,0,0,0,0,0,0,0,0,0,0,0\}\}$$

$\vec{J}$ orthogonal to each of these null eigenvectors means it is conserved across each set of four plaquettes associated with ($h_{01}$, $h_{02}$, $h_{03}$, $h_{12}$, $h_{13}$, $h_{23}$). We point out that this null space structure exists for the gauge equivalent form[50] $L = -\partial_\lambda h_{\alpha\beta} \partial^\lambda h^{\alpha\beta}$ on the lattice[12], but we included the mixing terms for comparison with the particular gauge choice in the photon case. We are now in position to interpret the Standard Model per MLGT, which makes clear how we approach unification and QG.

**4. Unification and Quantum Gravity**

*4.1 The Standard Model.* Strictly speaking, when finding the gradient of a vector field on the graph as we did with the Dirac operator, we need to specify a means of parallel transport. So, in our view and that of LGT, local gauge invariance is seen as a modification to the matter field gradient on the graph required by parallel transport per $U_\mu$, i.e.,

$$\gamma^\mu D_\mu \psi = \gamma^0 \left( \frac{U_0 \tilde{\psi}_0 - \psi}{ct} \right) + \gamma^1 \left( \frac{U_1 \tilde{\psi}_1 - \psi}{x} \right) + \ldots \tag{41}$$

where $\tilde{\psi}_i$ is the vector field on the node adjacent to $\psi$ in the positive $i^{\text{th}}$ direction. The Lagrangian density $L = \overline{\psi}\left(i\gamma^\mu \partial_\mu + e\gamma^\mu A_\mu - m\right)\psi - \frac{1}{4\mu_o} F^{\alpha\beta} F_{\alpha\beta}$ is therefore seen as the addition of parallel transport and a curvature term $A^\dagger \left(\partial_2{}^\dagger \partial_2\right) A$, where $A$ generates the parallel transport, to $L = \frac{1}{2}\overline{\psi}\left(\sqrt{\partial_1{}^\dagger \partial_1}\right)\psi$ to produce a well-defined field gradient between $\tilde{\psi}_i$ and $\psi$. $\vec{\vec{K}}$ now has the form $\vec{\vec{K}} = \begin{bmatrix} \begin{pmatrix} Dirac \\ plus \\ parallel \\ transport \end{pmatrix} & 0 \\ 0 & (Maxwell) \end{bmatrix}$ where Dirac $\vec{\vec{K}}$ has been modified to contain $A_\mu$

[12] The second term $2\partial_\lambda h^{\lambda\mu} \partial_\alpha h^\alpha_\mu$ of the gauge equivalent form would be used for multiple, connected graphical elements.

$$T=\begin{bmatrix} \frac{1}{t}+eA_0-m & 0 & 0 & 0 \\ 0 & \frac{1}{t}+eA_0 & 0 & 0 \\ 0 & 0 & \frac{1}{t}+eA_0 & 0 \\ 0 & 0 & 0 & \frac{1}{t}+eA_0 \end{bmatrix} \quad X=\begin{bmatrix} 0 & 0 & 0 & \frac{1}{x}+eA_1 \\ 0 & -m & \frac{1}{x}+eA_1 & 0 \\ 0 & \frac{1}{x}+eA_1 & 0 & 0 \\ \frac{1}{x}+eA_1 & 0 & 0 & 0 \end{bmatrix}$$

$$Y=\begin{bmatrix} 0 & 0 & 0 & -\frac{i}{y}-ieA_2 \\ 0 & 0 & \frac{i}{y}+ieA_2 & 0 \\ 0 & -\frac{i}{y}-ieA_2 & m & 0 \\ \frac{i}{y}+ieA_2 & 0 & 0 & 0 \end{bmatrix} \quad Z=\begin{bmatrix} 0 & 0 & \frac{1}{z}+eA_3 & 0 \\ 0 & 0 & 0 & -\frac{1}{z}-eA_3 \\ \frac{1}{z}+eA_3 & 0 & 0 & 0 \\ 0 & -\frac{1}{z}-eA_3 & 0 & m \end{bmatrix}$$

$A_\mu$ can have different values on different links, as required for non-zero $F_{\alpha\beta}$. But, even when $A_\mu$ has different values on different links, each row of the Dirac-plus-parallel-transport $\vec{\vec{K}}$ sums to zero, since it still has the form of Eq (37), so it possesses a non-trivial null space (that of Maxwell $\vec{\vec{K}}$ is obviously unaffected). The transition amplitude no longer has the simple Gaussian form since Dirac $\vec{\vec{K}}$ is now a function of one of the fields of integration. And $\vec{J}$ in the row space of $\vec{\vec{K}}$ now contains terms on links and nodes, representing conservation of 4-momentum between the interacting fields.

By introducing two vectors at each node, this same standard requires

$$\gamma^\mu D_\mu \psi=\gamma^0\left(\frac{\begin{bmatrix} C_{011} & C_{012} \\ C_{021} & C_{022}\end{bmatrix}\begin{bmatrix}\tilde{\psi}_0^1 \\ \tilde{\psi}_0^2\end{bmatrix}-\begin{bmatrix}\psi^1 \\ \psi^2\end{bmatrix}}{ct}\right)+\gamma^1\left(\frac{\begin{bmatrix} C_{111} & C_{112} \\ C_{121} & C_{122}\end{bmatrix}\begin{bmatrix}\tilde{\psi}_1^1 \\ \tilde{\psi}_1^2\end{bmatrix}-\begin{bmatrix}\psi^1 \\ \psi^2\end{bmatrix}}{x}\right)+\dots \qquad (42)$$

where the matrix $C_{\mu ab}$ is an element of SU(2) associated with the link in the positive $\mu^{th}$ direction from $\begin{bmatrix}\psi^1 \\ \psi^2\end{bmatrix}$. Again, we have the same form for our field gradients, i.e., the nodal field gradients parallel transported by the link field, which still contributes a gradient to the Lagrangian density $-\frac{1}{4g^2}\sum_{a,\alpha,\beta} F^a_{\alpha\beta}F^a_{\alpha\beta}$ where $g$ is the coupling constant and $a$ = 1, 2, 3, since SU(2) has three generators. Each link now has three different values for the gauge field, which we label $A^a_\mu$ with $a$ = 1, 2, 3. And, each of the three values can be different on different links. We have $F^a_{\alpha\beta} = \partial_\alpha A^a_\beta - \partial_\beta A^a_\alpha - g f^a_{bc} A^b_\alpha A^c_\beta$, where $f^a{}_{bc}$ are the SU(2) structure constants, $i\varepsilon^a{}_{bc}$. So, for example, Eq (39) modified for SU(2) gives $F^1_{01} = \frac{\tilde{A}^1_1 - A^1_1}{t} - \frac{\tilde{A}^1_0 - A^1_0}{x} - igA^2_0A^3_1 + igA^3_0A^2_1 - ig\tilde{A}^2_0\tilde{A}^3_1 + ig\tilde{A}^3_0\tilde{A}^2_1$, and you can see that now the pure gauge part ("Maxwell" part) of the Lagrangian density contains third and fourth-order terms in the gauge field. Thus, Maxwell $\vec{\vec{K}}$ now contains the gauge field, just like Dirac-plus-parallel-transport $\vec{\vec{K}}$. We can symmetrize this Maxwell+ $\vec{\vec{K}}$ so that the rows sum to zero and it possesses a non-trivial null space.

The Dirac-plus-parallel-transport $\vec{\vec{K}}$ is obtained using the 2x2 matrix-valued $A_\mu$ associated with each link given by $A_\mu = \sum_{a=1}^{3} A^a_\mu \frac{\sigma^a}{2}$ where $\sigma^a$ is a Pauli matrix. So, for example, $T$ is now an 8x8 matrix and the (1,1) entry of its 4x4 U(1) form becomes

$$\frac{1}{t} + eA_0 - m \rightarrow \begin{bmatrix} \frac{1}{t} + \frac{gA^3_0}{2} - m & \frac{gA^1_0 - igA^2_0}{2} \\ \frac{gA^1_0 + igA^2_0}{2} & \frac{1}{t} - \frac{gA^3_0}{2} - m \end{bmatrix}$$

As with U(1), each row of the Dirac-plus-parallel-transport $\vec{\vec{K}}$ sums to zero, since it still has the form of Eq (37), so it and Maxwell+ $\vec{\vec{K}}$ possess a non-trivial null space. It is now the case that both the matter field and gauge field portions of $\vec{\vec{K}}$ contain the gauge field. Thus, we see the progression from free field $\vec{\vec{K}}$ to abelian-interaction $\vec{\vec{K}}$ to non-abelian-interaction $\vec{\vec{K}}$ is a simple progression from $\vec{\vec{K}}$ with no gauge field terms to Dirac $\vec{\vec{K}}$ with gauge field terms to both Dirac and Maxwell $\vec{\vec{K}}$ with gauge field terms.

This pattern contains SU(3) where each link has eight different values for the gauge field (one for each generator of SU(3)) which we label $A_\mu^a$ with $a = 1, 2, \ldots, 8$. And, each of the eight values can be different on different links. As with SU(2), we have $F_{\alpha\beta}^a = \partial_\alpha A_\beta^a - \partial_\beta A_\alpha^a - g f_{bc}^a A_\alpha^b A_\beta^c$, where $f^a{}_{bc}$ are now the SU(3) structure constants[13]. Again, the pure gauge part ("Maxwell" part) of the Lagrangian density contains third and fourth-order terms in the gauge field and we can symmetrize Maxwell+$\vec{\vec{K}}$ such that the rows sum to zero and it possesses a non-trivial null space.

The 3x3 matrix-valued $A_\mu$ associated with each link is given by $A_\mu = \sum_{a=1}^{8} A_\mu^a \frac{\lambda^a}{2}$ where $\lambda^a$ is a Gell-Mann matrix[14]. So, for example, $T$ is now a 12x12 matrix and the (1,1) entry of its 4x4 U(1) form becomes

$$\frac{1}{t} + eA_0 - m \rightarrow \begin{bmatrix} \frac{1}{t} + \frac{gA_0^3}{2} + \frac{gA_0^8}{2\sqrt{3}} - m & \frac{gA_0^1 - igA_0^2}{2} & \frac{gA_0^4 - igA_0^5}{2} \\ \frac{gA_0^1 + igA_0^2}{2} & \frac{1}{t} - \frac{gA_0^3}{2} + \frac{gA_0^8}{2\sqrt{3}} - m & \frac{gA_0^6 - igA_0^7}{2} \\ \frac{gA_0^4 + igA_0^5}{2} & \frac{gA_0^6 + igA_0^7}{2} & \frac{1}{t} - \frac{gA_0^8}{\sqrt{3}} - m \end{bmatrix}$$

As with SU(2) and U(1), each row of the Dirac-plus-parallel-transport $\vec{\vec{K}}$ sums to zero, so it and Maxwell+$\vec{\vec{K}}$ possess a non-trivial null space. All possible mixing between U(1), SU(2), and SU(3) forms the Standard Model. We next explain particle physics per RBW.

*4.2 Particle Physics*. In our graphical approach, the role of the field is different than in the continuum approach of QFT where it pervades otherwise empty, continuous space to mediate the exchange of matter-energy between sources. In MLGT, as in LGT, the field is a scalar or vector associated with nodes, links, or plaquettes on the graph. One obtains

---

[13] In fact, this pattern extends to SU(N).

[14] This also generalizes to SU(N).

QFT results from LGT by letting the lattice spacing go to zero. In fact, one can understand QFT renormalization through this process of lattice regularization(51). As it turns out, however, this limit does not always exist, so calculated values are necessarily obtained from small, but non-zero, lattice spacing(52). With this picture in mind, we can say simply what we are proposing: The lattice is fundamental, not its continuum limit. Once one accepts this premise, it's merely a matter of degree to have large spacetimesource elements, which is the basis for our explanation of the twin-slit experiment (section 3.4 above) and dark energy (section 5.2 below). In this approach, *there is no graphical counterpart to "quantum systems" traveling through space as a function of time from Source to sink to "cause" detector clicks.* This implies the empirical goal at the fundamental level is to tell a unified story about detector events to include individual clicks – how they are distributed in space (e.g., interference patterns, interferometer outcomes, spin measurements), how they are distributed in time (e.g., click rates, coincidence counts), how they are distributed in space and time (e.g., particle trajectories), and how they generate more complex phenomena (e.g., photoelectric effect, superconductivity). Thus in our theory, particle physics per QFT is in the business of characterizing large sets of detector data, i.e., all the individual clicks.

As was eminently apparent from our examples in section 3, it is practically impossible to compute *Z*(*J*) in MLGT for all possible spatiotemporally relative click locations in a particle physics "event," which contains "approximately 100,000 individual measurements of either energy or spatial information(53)." However, we know from theory(54) and experiment that, with overwhelming probability, detector clicks will trace classical paths[15], so it makes sense to partition large click distributions into individual trajectories and treat these as the fundamental constituents of high energy physics experiments[16]. This is exactly what QFT does for particle physics according to our interpretation. Since the individual trajectories are themselves continuous, QFT uses propagators in continuous spacetime which entails an indenumerably infinite number of locations for both clicks and interaction vertices. *Thus, issues of regularization and renormalization are simply consequences of the continuum approximation necessary to deal with very large click distributions, having decided to parse the click distributions into continuum trajectories*.

[15] Individual detector clicks (called "hits in the tracking chamber") are first localized spatially (called "preprocessing"), then associated with a particular track (called "pattern recognition"). The tracks must then be parameterized to obtain dynamical characteristics (called "geometrical fitting"). See Fernow, R.C.: Introduction to experimental particle physics. Cambridge University Press, Cambridge (1986), sections 1.7.1, 1.7.2 & 1.7.3, respectively.

Essentially, we're saying a particle physics detector event is one giant interference pattern, as we previously characterized interference, and the way to understand a particular pattern involving thousands of clicks can only realistically be accomplished by parsing an event into smaller subsets, and the choice of subsets is empirically obvious, i.e., spacetime trajectories. These trajectories are then characterized by mass, spin, and charge. Per RBW's adynamical explanation, the colliding beams in the accelerator and the detector surrounding the collision point form the graphical input that, in conjunction with the AGC, dictate the possible configurations of spacetimesource elements responsible for particle trajectories. Each trajectory can be deduced one click at a time in succession using $\psi(x,t)$, as shown by Mott for alpha particles in a cloud chamber[(55)]. Therefore, a probability amplitude could be computed for each trajectory using spacetimesource elements detection event by detection event *a la* our twin-slit analysis above. However, as shown by Mott, after the first click the remaining clicks follow a classical trajectory with high probability, so the only real quantum computation needed is for the probability amplitude of the spacetimesource element of the set of first clicks, i.e., the first click for each trajectory in the collection. And, the properties (mass, charge, momentum, energy, etc.) for that spacetimesource element would simply be the properties of the subsequent particles defined relationally in the context of the accelerator Source and particle detector. In standard LGT → QFT the calculated outcomes are found by taking the limit as the lattice spacing goes to zero via renormalization, but we needn't assume the spacing goes to zero, since our sinks are the pixel locations in the detector CCD. Likewise, assuming the accelerator and detectors are sufficiently isolated during the brief period of data collection, the graph size is not infinite as in QFT. This of course justifies the UV and IR cutoffs in regularization, respectively.

---

[16] Some assumptions are required, e.g., "Sometimes it is necessary to know the identity (i.e., the mass) of at least some of the particles resulting from an interaction" (Fernow, 1986, p 17), "Within the errors [for track measurements], tracks may appear to come from more than one vertex. Thus, the physics questions under study may influence how the tracks are assigned to vertices" (Fernow, 1986, p 25), and "Now there must be some minimum requirements for what constitutes a track. Chambers may have spurious noise hits, while the chambers closest to the target may have many closely spaced hits. The position of each hit is only known to the accuracy of the chamber resolution. This makes it difficult to determine whether possible short track combinations are really tracks" (Fernow, 1986, p 22). Despite these assumptions, no one disputes the inference. While we do not subscribe to the existence of "click-causing entities" with trajectories, we agree that clicks trace classical paths. Indeed, this is the basis for our approach and consequently, the results and analysis of particle physics experiments are very important.

This severely undermines the dynamical picture of perturbations moving through a continuum medium (naïve field) between sources, i.e., it undermines the naïve notion of a particle as traditionally understood. In fact, the typical notion of a particle is associated with the global particle state of n-particle Fock space and per Colosi & Rovelli "the notion of global particle state is ambiguous, ill-defined, or completely impossible to define[(56)]." What we mean by "particle" is a collection of detector hits forming a spacetime trajectory resulting from a collection of adynamically constrained spacetimesource elements in the presence of colliding beams and a detector. And this doesn't entail the existence of an object with intrinsic properties, such as mass and charge, moving through the detector to cause the hits.

Our view of particles agrees with Colosi & Rovelli[(57)] on two important counts. First, that particles are best modeled by local particle states rather than n-particle Fock states computed over infinite regions, squaring with the fact that particle detectors are finite in size and experiments are finite in time. The advantage to this approach is that one can unambiguously define the notion of particles in curved spacetime as excitations in a local M4 region, which makes it amenable to Regge calculus. Second, this theory of particles is much more compatible with the quantum notion of complementary observables in that every detector has its own Hamiltonian (different sized graph), and therefore its own particle basis (unlike the unique basis of Fock space). Per Colosi & Rovelli, "In other words, we are in a genuine quantum mechanical situation in which distinct particle numbers are complementary observables. Different bases that diagonalize different $H_R$ [Hamiltonian] operators have equal footing. Whether a particle exists or not depends on what I decide to measure." Thus, in our view, particles simply describe how detectors and Sources are relationally co-defined via the AGC.

*4.3 Unification and Quantum Gravity*. With this understanding of the Standard Model and particle physics, we see that the next logical addition to our collection of fundamental spacetimesource elements would be those constructed from the gradient of vector fields on links. The scalar field on plaquettes (basis for graviton in standard LGT) would define parallel transport for this field gradient in the manner scalar fields on links defines parallel transport for the vector fields on nodes. This is the standard approach to QG in the particle physics community. The problem with this is, of course, we simply have gravitons in M4, we still need spacetime curvature as in GR. In our view, since MLGT is necessarily contextual, that is accomplished by understanding the context of the properties in question *a la* the simplices of Regge calculus (see section 5.1 and Figure 13).

According to Regge calculus, gravity is a scalar field on the plaquettes of its simplices, i.e., Newtonian gravity in M4, and spacetime curvature (variable spacetime geometry) is accounted for via "deficit angles" between simplices in the global structure (Figure 13). This spacetime curvature is a function of the energy-momentum content of spacetime (section 5.1) to include all forms, not just the graviton (associated with the property of mass). Thus, if we were able to construct experiments with individual gravitons, a QG experiment could be the simple twin-slit experiment with gravitons in M4, i.e., in a single simplex. Or, we could view graviton interference patterns generated between a Source and detector in non-neighboring simplices, in which case the spacetimesource element for the exchange of gravitational energy would have simplex-to-simplex segments between Source and detection event. To compute the amplitude for the exchange of gravitational energy associated with that spacetimesource element, one would pick up phase factors associated with the deficit angles between simplices, just as a photon amplitude picks up phase factors associated with reflection from mirrors and beam splitters as computed between Source and detection event in an interferometer.

Since we don't yet have the technology to manipulate individual gravitons, we might rather explore the effect of variable geometry on spacetimesource elements passing through neighboring simplices for the exchange of photons (as on astrophysical scales). Since different energy distributions mean different spacetime geometries per Regge calculus, an exchange of energy, quantum or otherwise, has geometric implications. And, since geometry affects the value of the action in Regge calculus, one should expect that energy lost at the Source will not only reappear at the detection event, but will also appear in the spacetime geometry, as compared to the geometry where the photon energy exchange is not taken into account. We consider just such an example of this "disordered locality" in section 5.2 below. There the energy lost by the photon per its cosmological redshift between Source (supernova) and detector (telescope) in a cosmology model based on matter alone (no electromagnetic contribution to the spacetime geometry) will be used to justify a first-order correction to the proper distance between Source and detector in that dust-filled cosmology model. That correction is small (scaled by a factor of $(8.38\ \text{Gcy})^{-1}$), but we show that it does allow for a fit of distance modulus versus redshift for the Union2 Compilation supernovae data without accelerating expansion and, therefore, without dark energy.

At this point the reader should appreciate that underwriting interacting COs via spacetimesource elements leads to a relatively simple picture of unification (Figure 11) compared to that based on fundamental particles (Figure 12). However, while we do not view particle physics as the study of what is ultimately fundamental in Nature, it has been essential to understanding how the fundamental elements of spacetimesource are to be constructed and combined, and what properties are represented by $\bar{J}$, as we explained in sections 3.4 and 4.2. Since high energy particle physics deals with large energy densities, our disordered locality does not need to be taken into account. Disordered locality becomes a concern for relatively small energy exchanges over relatively large spatiotemporal regions. We encounter that situation in astrophysics, so that is where we expect disordered locality per our version of QG to become important.

## 5. Implications for Astrophysics and Cosmology

*5.1 Regge Calculus*. In Regge calculus, the spacetime manifold is replaced by a lattice geometry where each 4D cell (simplex) is Minkowskian (flat). Curvature is represented by "deficit angles" (Figure 13) about any plane orthogonal to a "hinge" (triangular side to a tetrahedron, which is a 3D side of a 4D simplex). The Hilbert action for a 4D vacuum lattice is $I_R = \frac{1}{8\pi}\sum_{\sigma_i \in L} \varepsilon_i A_i$ where $\sigma_i$ is a triangular hinge in the lattice $L$, $A_i$ is the area of $\sigma_i$ and $\varepsilon_i$ is the deficit angle associated with $\sigma_i$. The counterpart to Einstein's equations is then obtained by demanding $\frac{\delta I_R}{\delta \ell_j^2} = 0$, where $\ell_j^2$ is the squared length of the $j^{th}$ lattice edge , i.e., the metric. To obtain equations in the presence of matter-energy, one simply adds the appropriate term $I_{M\text{-}E}$ to $I_R$ and carries out the variation as before to obtain $\frac{\delta I_R}{\delta \ell_j^2} = -\frac{\delta I_{M-E}}{\delta \ell_j^2}$. One finds the stress-energy tensor is associated with lattice edges, just as the metric, and Regge's equations are to be satisfied for any particular choice of the two tensors on the lattice. [Notice, the adynamical global constraint nature of GR is particularly evident in the Regge calculus approach.]

*5.2 Dark Energy and Other Astrophysical Implications*. Since one recovers GR from Regge calculus by making the simplices small (as in LGT → QFT), it seems that empirical evidence of the deviation from GR phenomena posed by large spacetimesource elements of disordered locality, i.e., modified Regge calculus (MORC), might be found in the exchange of photons on cosmological scales. Therefore, we modified the Regge calculus approach to Einstein-deSitter cosmology (EdS)[58] and compared this MORC model, EdS, and the concordance model ΛCDM (EdS plus a cosmological constant Λ to account for dark energy) using the data from the Union2 Compilation, i.e., distance moduli and redshifts for type Ia supernovae[59] (Figure 14). We found that a best fit line through log($D_L$/Gpc) versus log($z$) gives a correlation of 0.9955 and a sum of squares error (SSE) of 1.95. By comparison, the best fit ΛCDM gives SSE = 1.79 using a Hubble constant of Ho = 69.2 km/s/Mpc, $\Omega_M = 0.29$ and $\Omega_\Lambda = 0.71$. The parameters for ΛCDM yielding the most robust fit to[60] "the Wilkinson Microwave Anisotropy Probe data with the latest distance measurements from the Baryon Acoustic Oscillations in the distribution of galaxies and the Hubble constant measurement" are Ho = 70.3 km/s/Mpc, $\Omega_M = 0.27$ and $\Omega_\Lambda = 0.73$, which are consistent with the parameters we find for its Union2 Compilation fit. The best fit EdS gives SSE = 2.68 using Ho = 60.9 km/s/Mpc. The best fit MORC gives SSE = 1.77 and Ho = 73.9 km/s/Mpc with the EdS proper distance $D_p$ corrected by a factor of $\sqrt{1+\frac{D_p}{A}}$ where $A = 8.38$ Gcy. A current "best estimate" for the Hubble constant is Ho = (73.8 ± 2.4*)* km/s/Mpc[61]. Thus, MORC improves EdS as much as ΛCDM in accounting for distance moduli and redshifts for type Ia supernovae even though the MORC universe contains no dark energy is therefore always decelerating. So, per MLGT, it is quite possible that this data does not constitute "the discovery of the accelerating expansion of the Universe," (Nobel citation, 2011), i.e., there is no accelerating expansion, so there is no need of a cosmological constant or dark energy in any form[62].

Our theory has other possible implications for astrophysics and cosmology as well. Perhaps MORC's version of the Schwarzschild solution will negate the need for dark matter as its counterpart to Einstein-deSitter cosmology did with dark energy. What will MORC have to say about the event horizon and singularity in the Schwarzschild solution, i.e., black holes? Perhaps, the singularity will be avoided as in Regge calculus cosmology where backwards time evolution "stops" at a time determined by the choice of lattice spacing[17]. And, with an adynamical approach, cosmological explanation takes on an entirely new form. No longer is one seeking explanation in the form of a time-evolved spatial hypersurface of homogeneity – an explanation that cannot be satisfied with the Big Bang or even a non-singular "stop point." Thus, such dynamical explanation results in contentious, misleading or unverifiable notions about[63] "creation from nothing," the multiverse, etc. Rather, explanation via adynamical self-consistency writ large doesn't rest ultimately on the Big Bang or any other region of the graph[18]. The reason the fields on node X and link Y have the values they do is required by the solution for the entire graph, i.e., it is required by the values of the fields on all the other nodes and links. As we pointed out in section 1 when we contrasted dynamical explanation with our adynamical global constraint explanation, no region of the graph is distinguished over any other in this explanatory scheme.

[17] This is the "stop point problem" of Regge calculus cosmology. Of course it's not a "problem" for our approach, since Regge calculus is fundamental to GR, not the converse, one does not require Regge calculus reproduce the initial singularity of GR cosmology.

## 6. Conclusion

Our combination of MLGT and MORC is an entirely new way to reconcile QFT with GR and unify physics:

$$\text{QFT} \leftarrow \text{MLGT} \leftrightarrow \text{MORC} \rightarrow \text{GR}$$

Many of the major questions that need to be answered in this new view of unification and QG are clear. Is there a limit to the number of vectors that can be (or need be) introduced on nodes and links? If so, does it have to do with information density? Is it related to quark confinement? Or, is there a purely mathematical fact that underwrites it? Why is there no physical counterpart to a scalar field on cubes? Is this because it requires (4+1)D to close graphically and satisfy the boundary of a boundary principle for all graphical entities? What physical objects correspond to vector fields on links? Are they just quarks and leptons interacting gravitationally? Or, will this generate new fermions that only interact gravitationally, e.g., dark matter? While these questions are not going to be answered in a trivial fashion, we believe the RBW program of unification and QG offers a viable alternative to the existing landscape. In fact, MORC has already produced an empirical result, i.e., an explanation of dark energy, as we showed in section 5.

Our explanation of dark energy resulted from RBW's modification to GR cosmological proper distance per disordered locality (MORC). Astrophysical data is very amenable to analysis via MORC, since it represents low energy exchanges over large spatiotemporal extents. Thus, we expect that dark matter is also a candidate for explanation via kinematical corrections per MORC. In contrast, high energy physics deals with large energy densities and that is precisely where we expect analytic techniques such as those of the Standard Model to work well. Thus, we don't suggest any sweeping changes to the formalism of particle physics as it is currently employed. Rather, MLGT vindicates the formalism by providing rationale for some of its questionable techniques, e.g., UV and IR cutoffs in regularization. Instead, we would expect to see corrections to QFT in the low energy regime, which is where QM takes over. In fact, it is in this regime where experiments have vindicated some of QM's most "mysterious" predictions, e.g., delayed-choice experiments, and it was just such phenomena that inspired RBW. Thus, we expect to expand the list of "mysterious" quantum predictions in the context of QFT by looking for particle physics effects related to detector size and shape, for example. We will stop such speculation at this point, given the incipient nature of MLGT and MORC.

[18] See Rovelli, C.: Why do we remember the past and not the future? The 'time oriented coarse graining' hypothesis. http://arxiv.org/abs/1407.3384 for an idea that is similar in spirit.

**Figure 1**

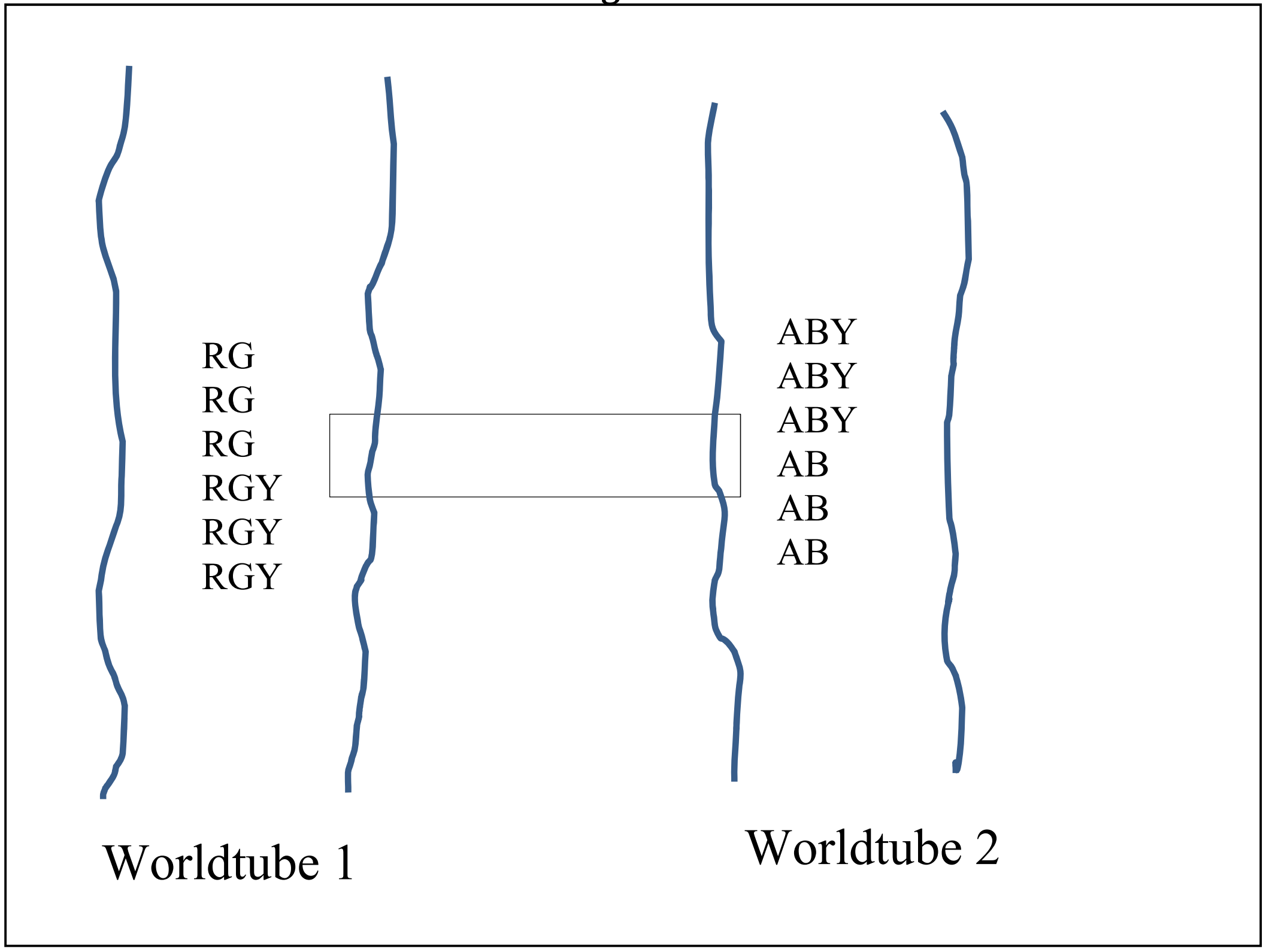

**Quantum Exchange of Energy-Momentum** – The property Y is associated with the source on the spacetimesource element (rectangle) shared by the worldtubes. As a result, property Y disappears from Worldtube 1 (Y Source) and reappears later at Worldtube 2 (Y detector) without mediation. That is, there is no third worldtube/line needed to explain the exchange of energy-momentum associated with property Y between Worldtube 1 and Worldtube 2. While these properties are depicted as residing in the worldtubes, they don't represent something truly intrinsic to the worldtubes, but are ultimately contextual and relational, i.e., being the Source of Y only makes sense in the context of (in relation to) a "Y detector", and vice-versa. The A, B, R, and G properties shown might be established with respect to classical objects not shown in this Figure, for example.

**Figure 2**

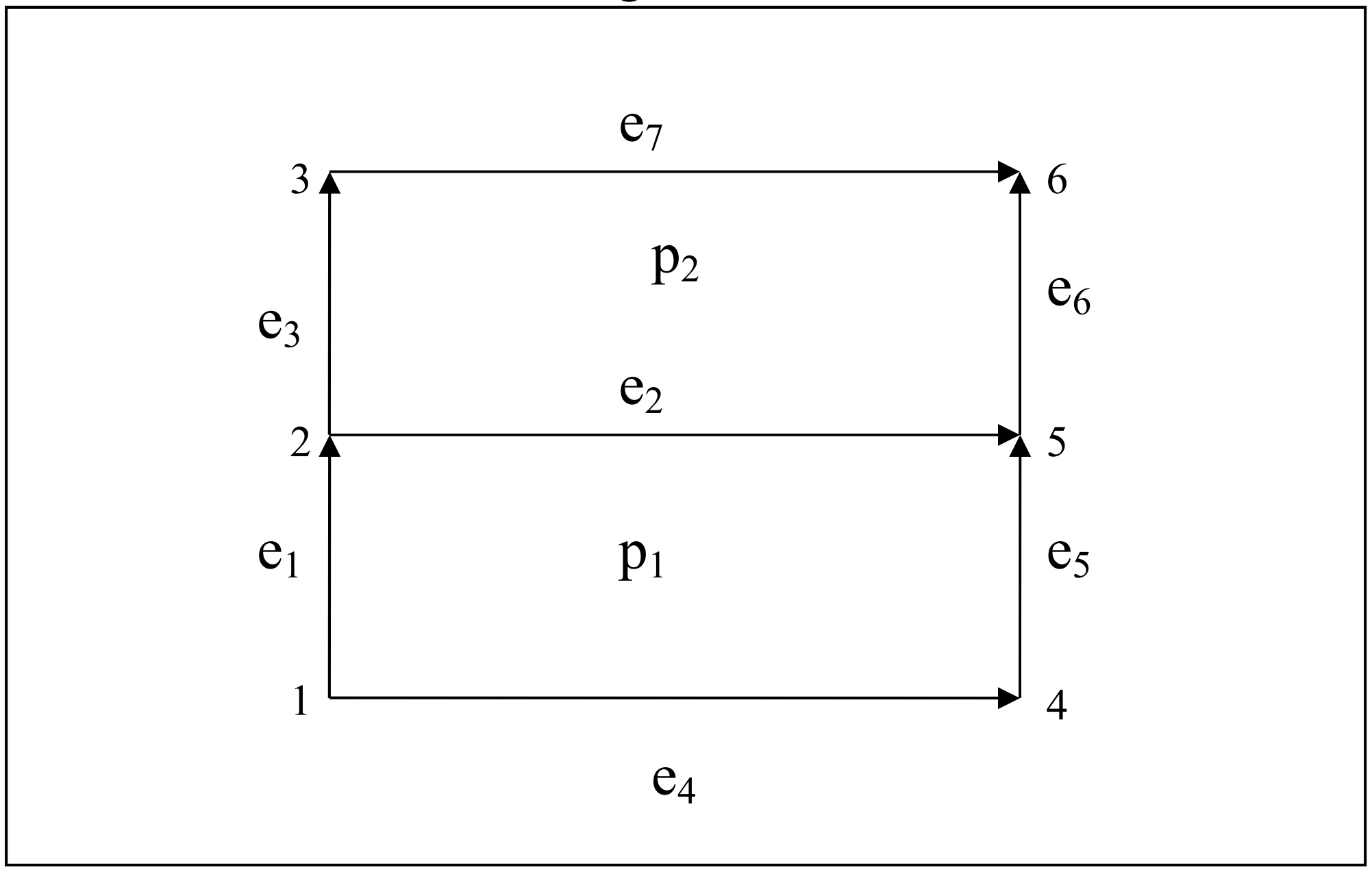

**Figure 3**

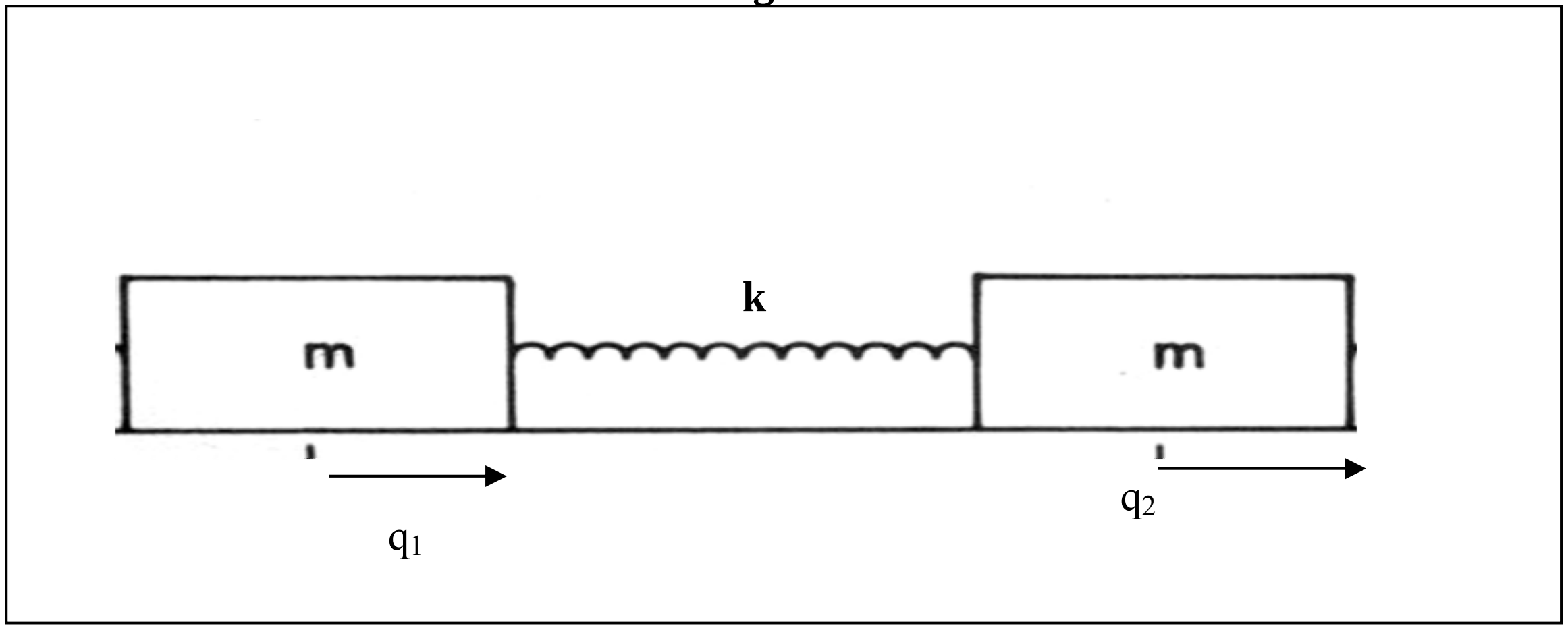

**Figure 4**

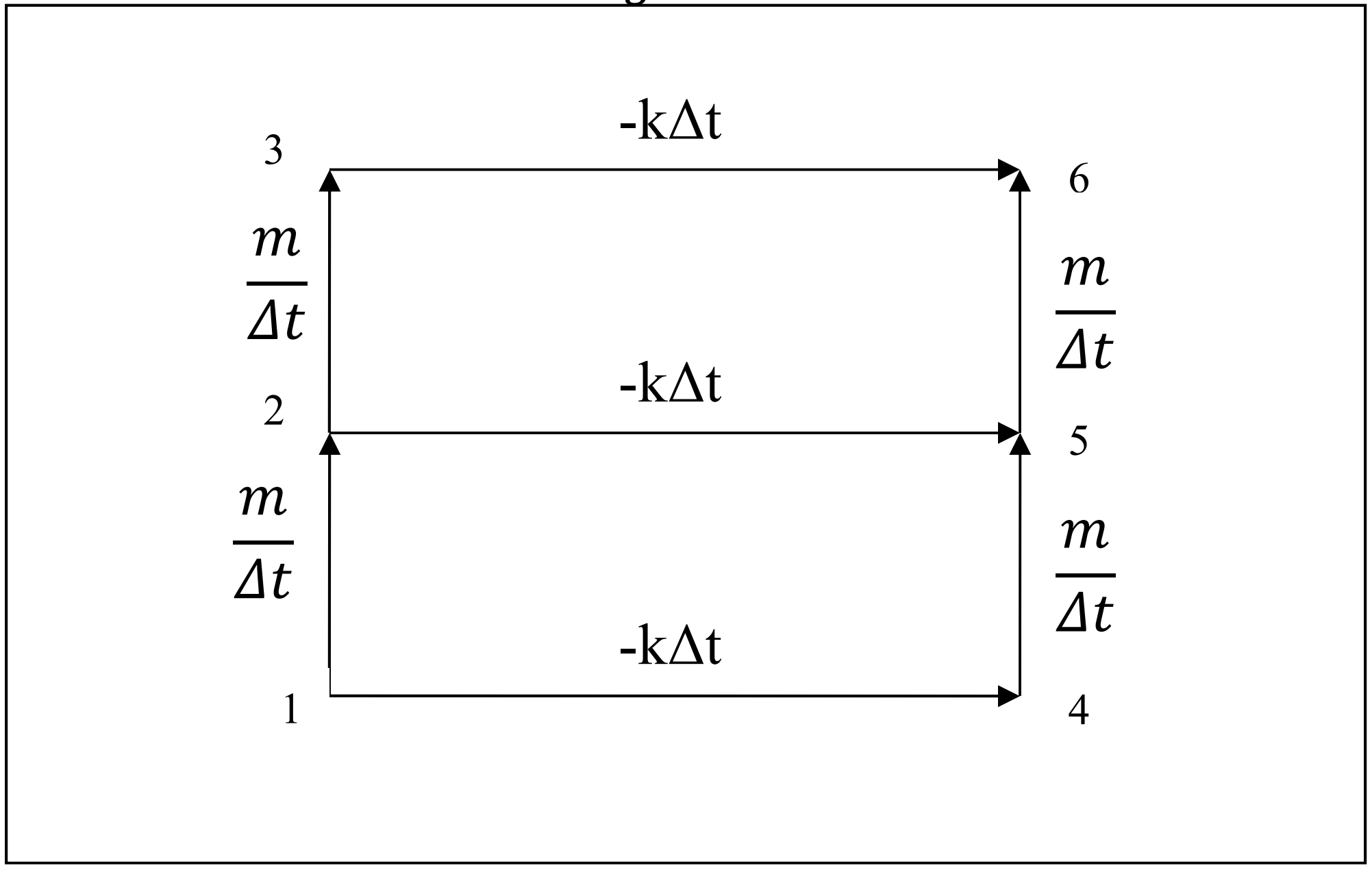

**Figure 5**

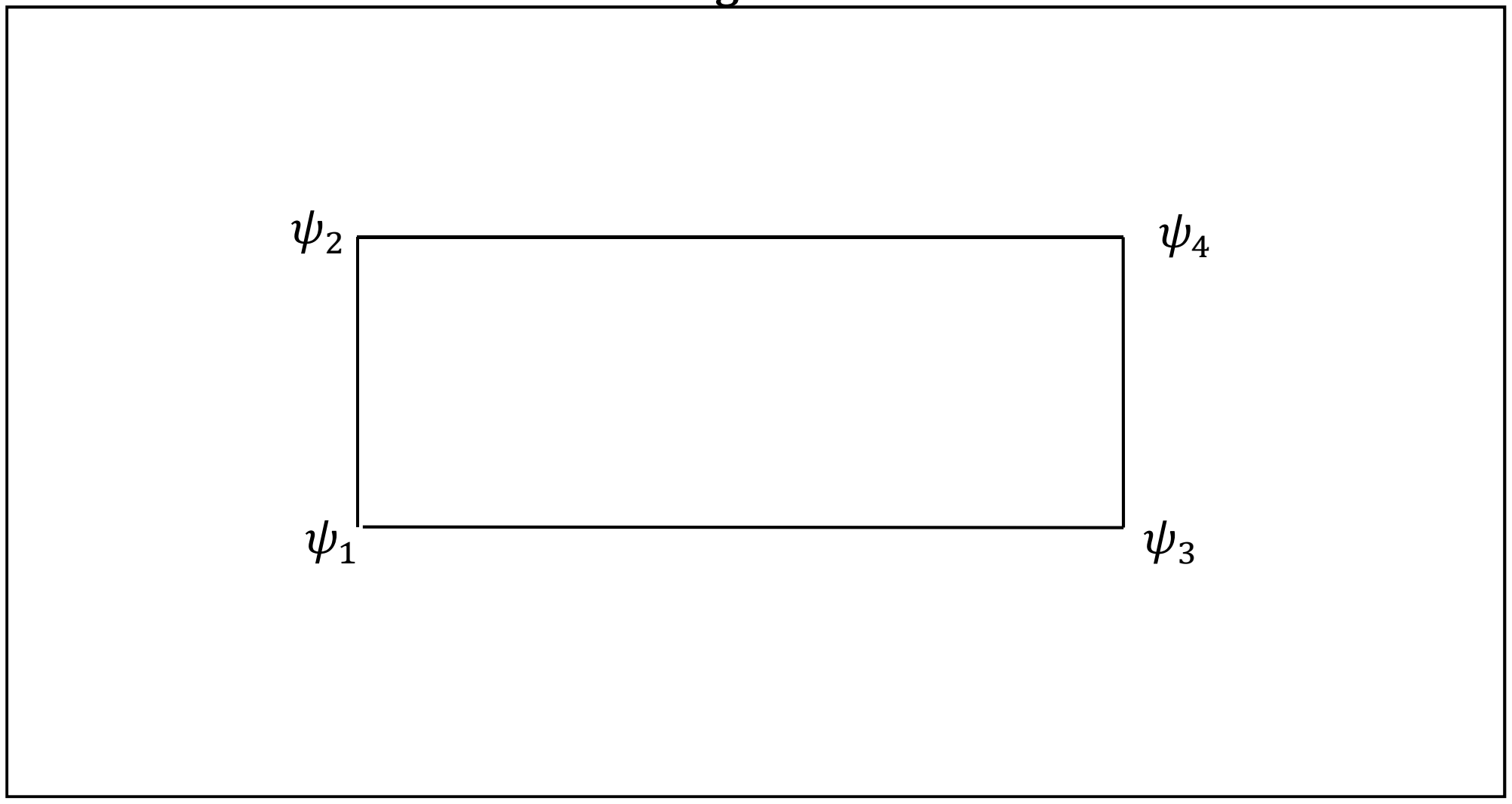

**Figure 6**
**Twin-Slit Interference**

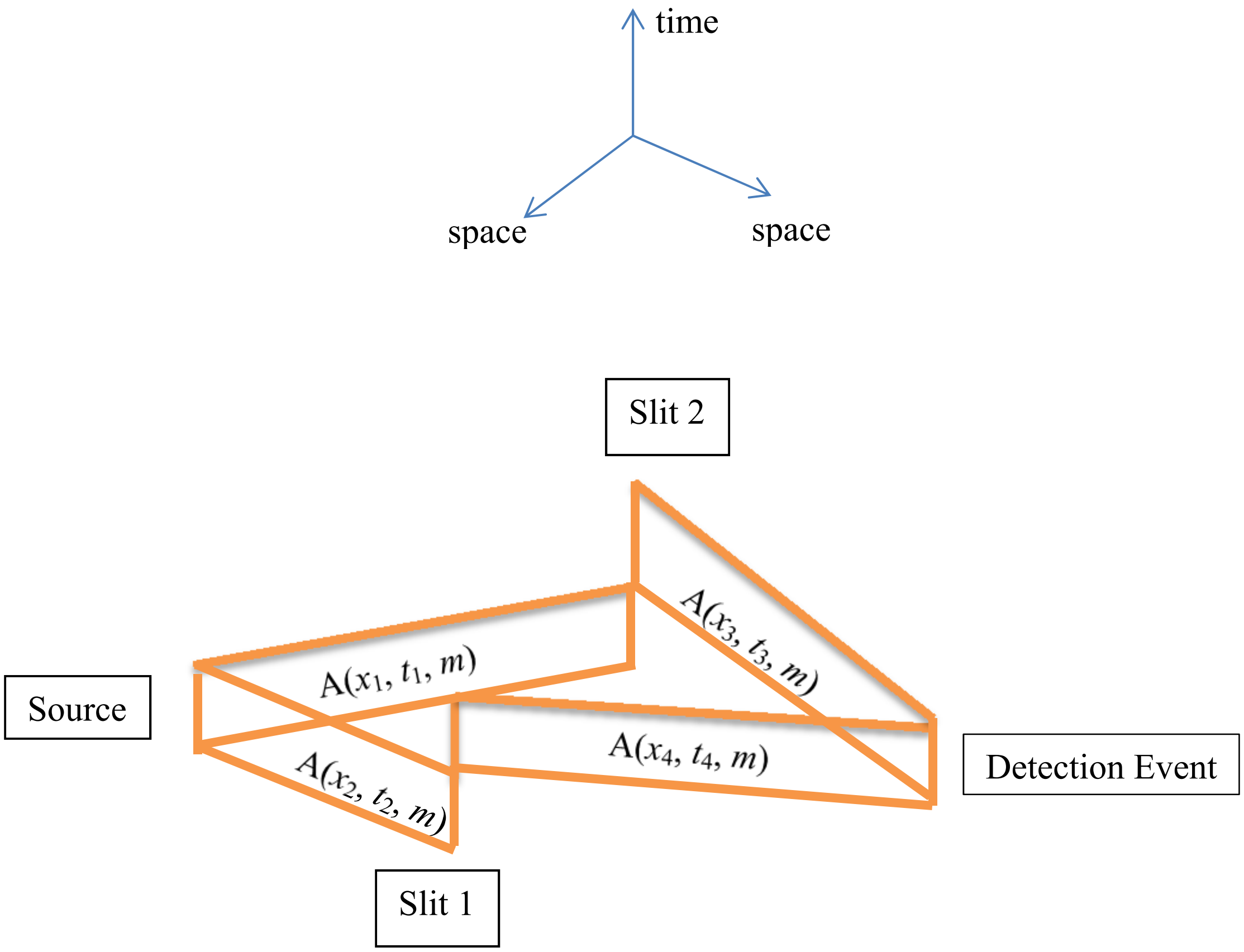

The boxes are the components of the spacetimesource element depicting mass *m* loss at the Source emission event and mass *m* gain at the Detection Event contributing to an interference pattern at the detector.

**Figure 7**

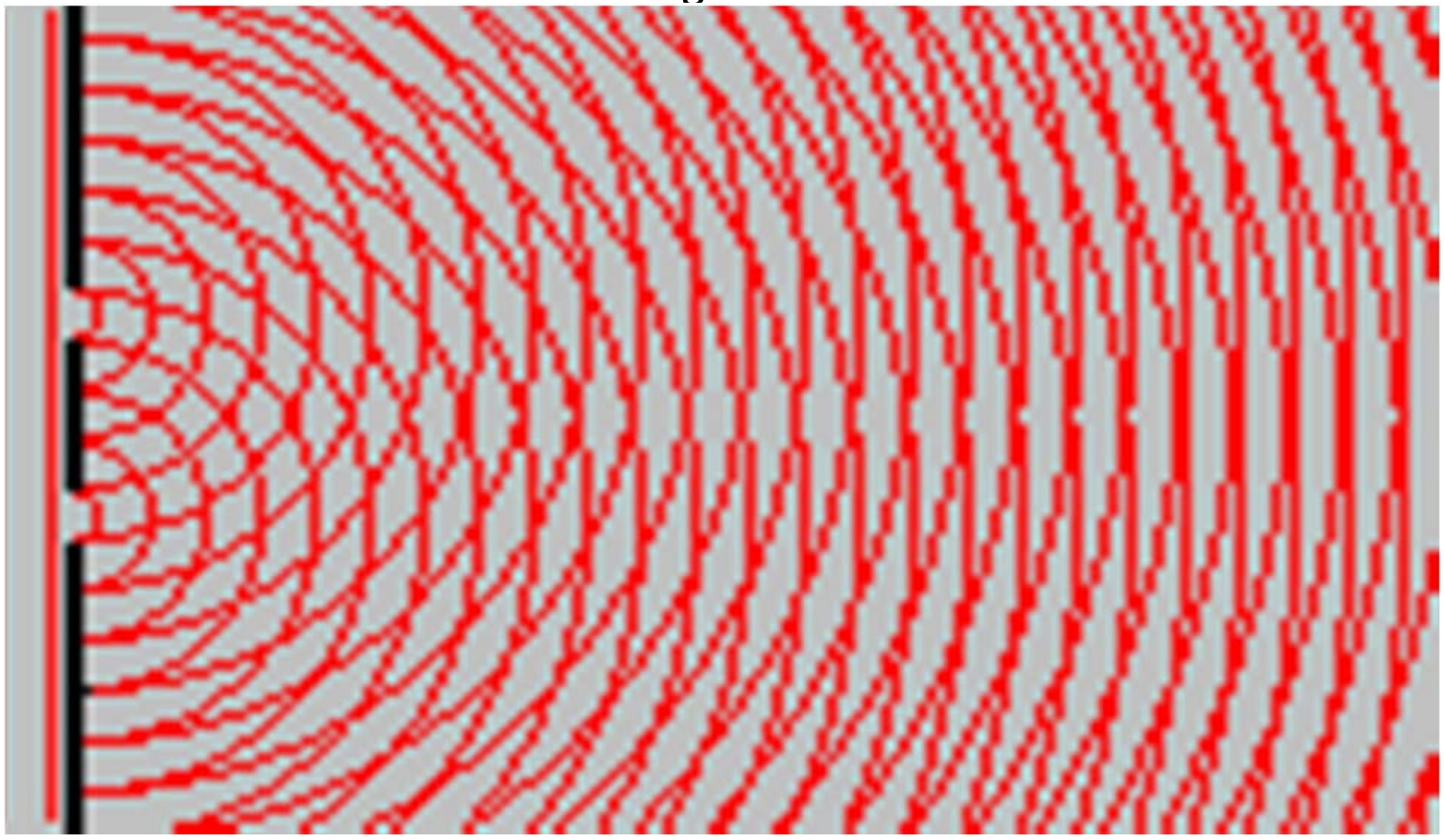

**Figure 8**

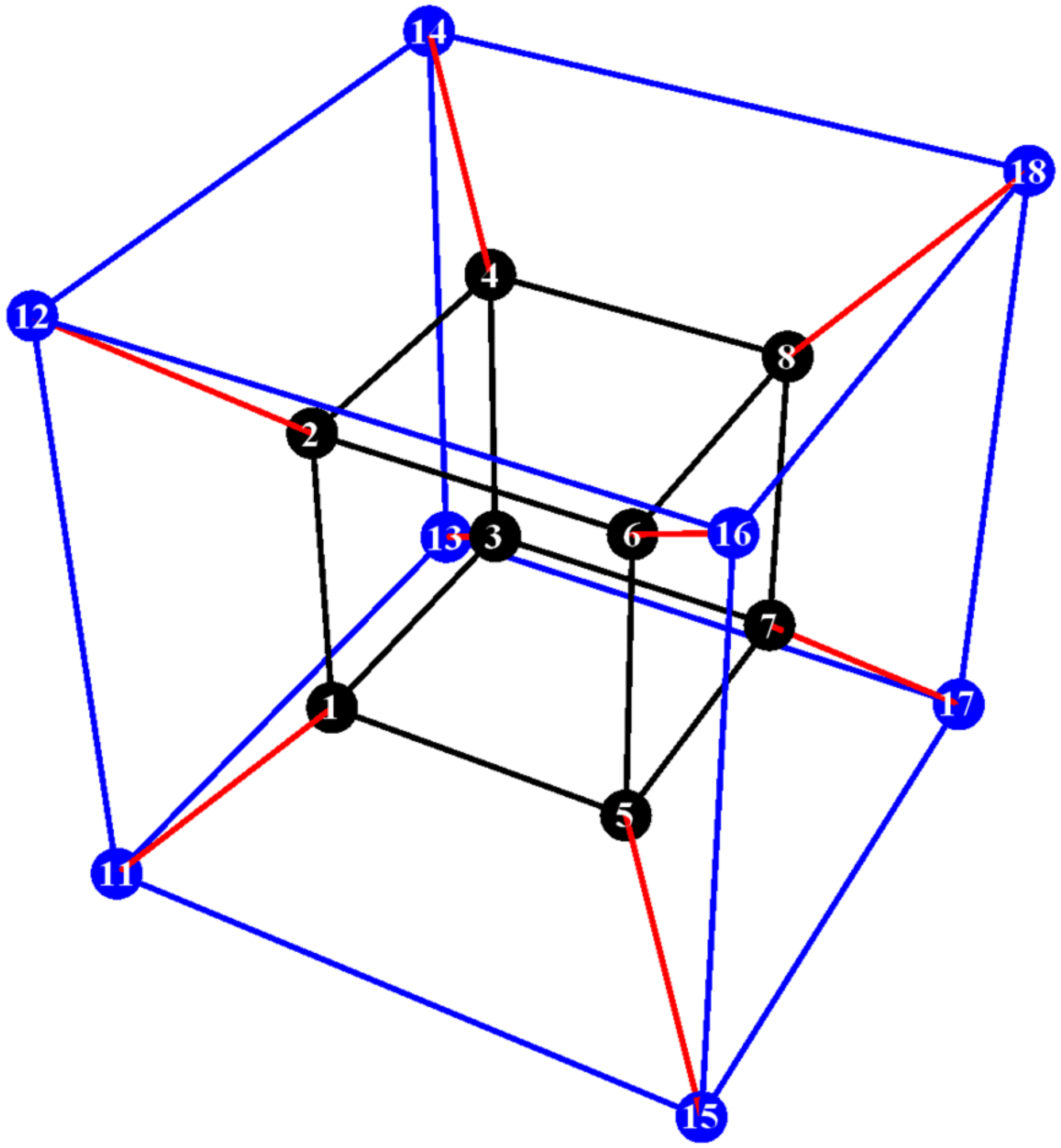

**Figure 9**

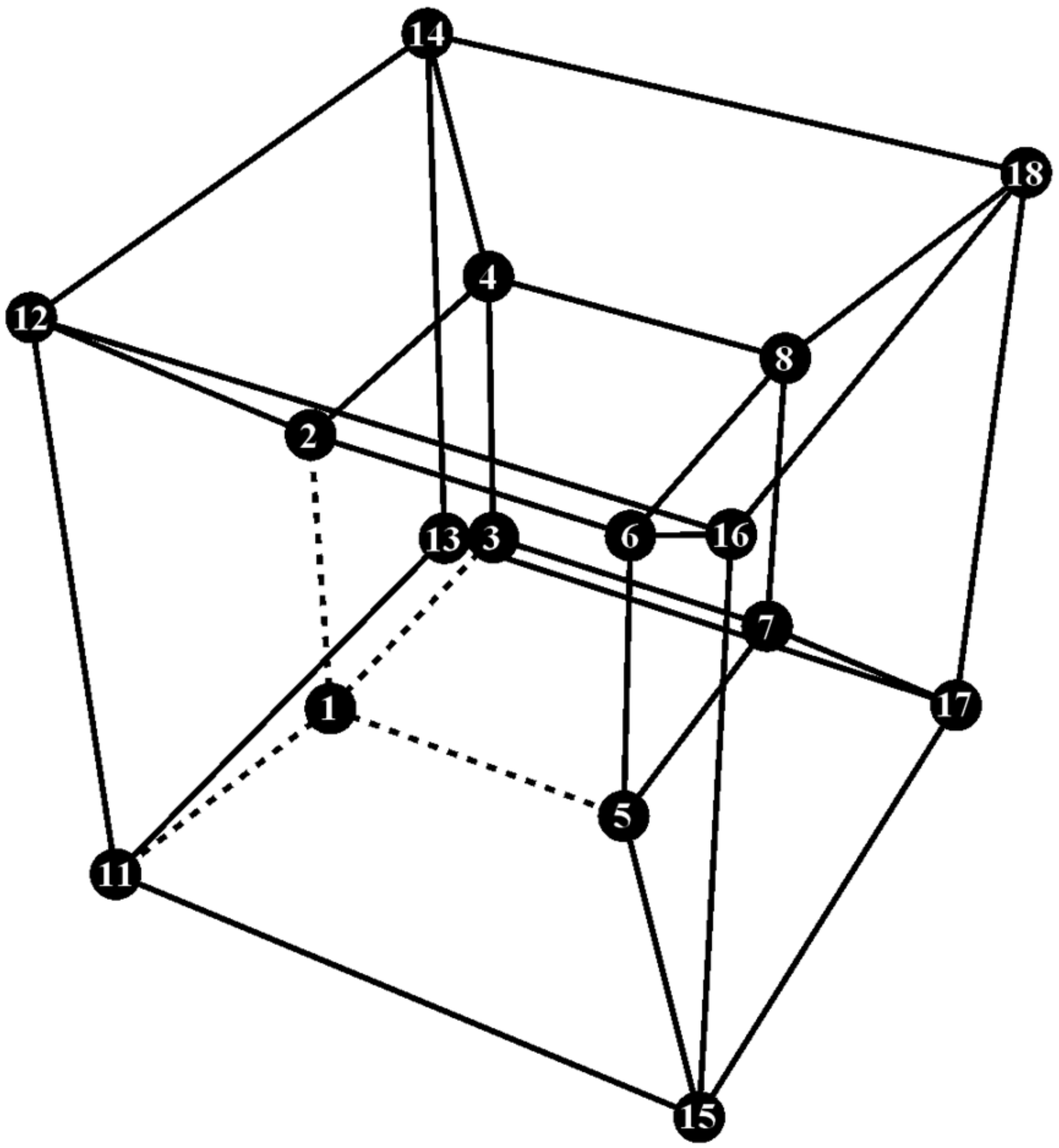

**Figure 10**

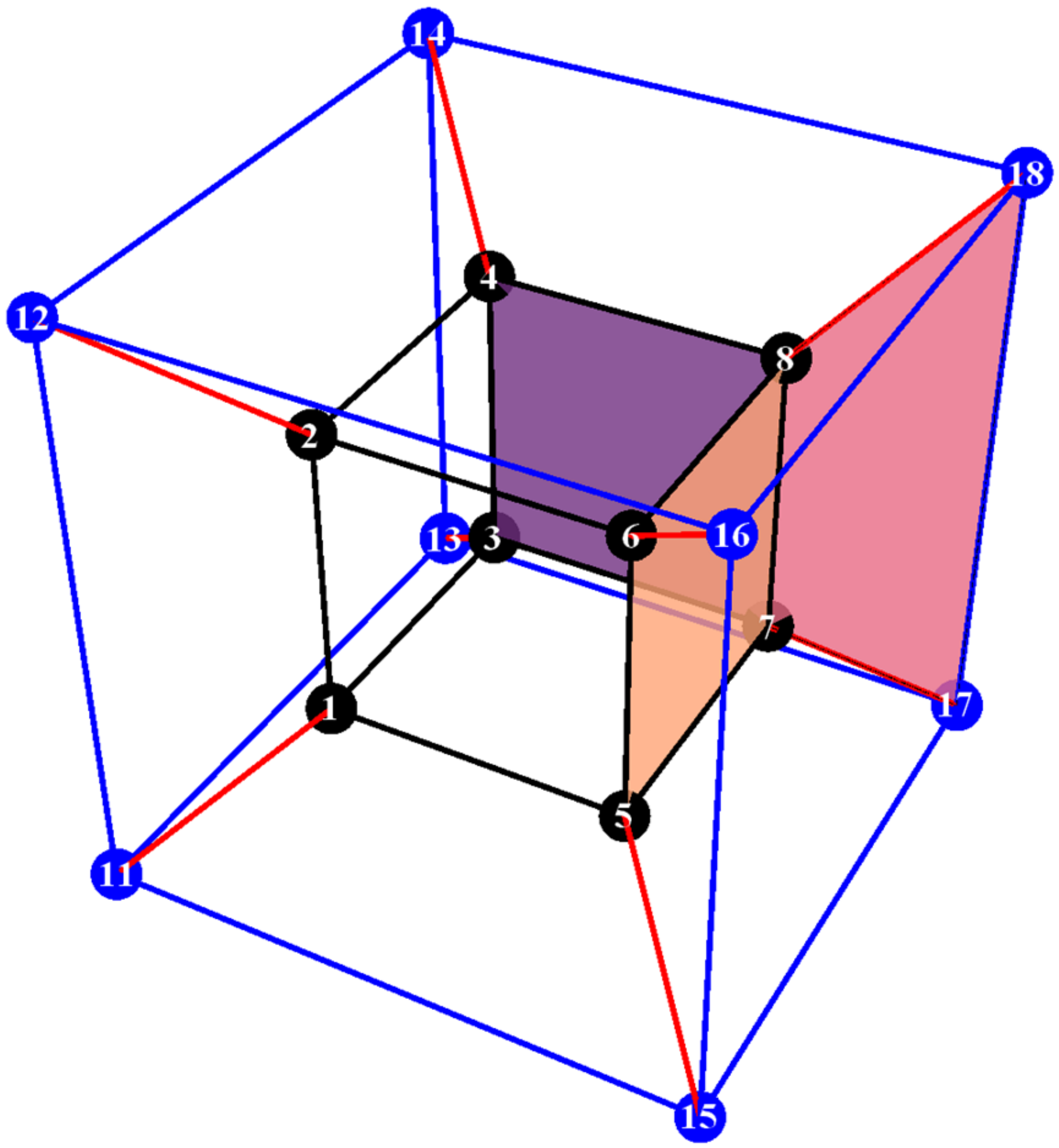

**Figure 11**

**Fundamental spacetimesource elements for unification via MLGT**

| | | |
|---|---|---|
| Scalar field on nodes | One vector each node | One vector each link |
| Scalar field on links | Two vectors each node | Two vectors each link |
| Scalar field on plaquettes | Three vectors each node | Three vectors each link |

**Figure 12**

**The Standard Model Lagrangian Density. Credit: T.D. Gutierrez**

http://nuclear.ucdavis.edu/~tgutierr/files/sml.pdf

$$
\begin{aligned}
&-\tfrac12\partial_\nu g^a_\mu\partial_\nu g^a_\mu - g_s f^{abc}\partial_\mu g^a_\nu g^b_\mu g^c_\nu - \tfrac14 g_s^2 f^{abc}f^{ade} g^b_\mu g^c_\nu g^d_\mu g^e_\nu + \\
&\tfrac12 ig_s^2(\bar q^\sigma_i\gamma^\mu q^\sigma_j)g^a_\mu + \bar G^a\partial^2 G^a + g_s f^{abc}\partial_\mu \bar G^a G^b g^c_\mu - \partial_\nu W^+_\mu\partial_\nu W^-_\mu - \\
&M^2 W^+_\mu W^-_\mu - \tfrac12\partial_\nu Z^0_\mu\partial_\nu Z^0_\mu - \tfrac{1}{2c_w^2}M^2 Z^0_\mu Z^0_\mu - \tfrac12\partial_\mu A_\nu\partial_\mu A_\nu - \tfrac12\partial_\mu H\partial_\mu H - \\
&\tfrac12 m_h^2 H^2 - \partial_\mu\phi^+\partial_\mu\phi^- - M^2\phi^+\phi^- - \tfrac12\partial_\mu\phi^0\partial_\mu\phi^0 - \tfrac{1}{2c_w^2}M\phi^0\phi^0 - \beta_h[\tfrac{2M^2}{g^2} + \\
&\tfrac{2M}{g}H + \tfrac12(H^2+\phi^0\phi^0+2\phi^+\phi^-)] + \tfrac{2M^4}{g^2}\alpha_h - igc_w[\partial_\nu Z^0_\mu(W^+_\mu W^-_\nu - \\
&W^+_\nu W^-_\mu) - Z^0_\nu(W^+_\mu\partial_\nu W^-_\mu - W^-_\mu\partial_\nu W^+_\mu) + Z^0_\mu(W^+_\nu\partial_\nu W^-_\mu - \\
&W^-_\nu\partial_\nu W^+_\mu)] - igs_w[\partial_\nu A_\mu(W^+_\mu W^-_\nu - W^+_\nu W^-_\mu) - A_\nu(W^+_\mu\partial_\nu W^-_\mu - \\
&W^-_\mu\partial_\nu W^+_\mu) + A_\mu(W^+_\nu\partial_\nu W^-_\mu - W^-_\nu\partial_\nu W^+_\mu)] - \tfrac12 g^2 W^+_\mu W^-_\mu W^+_\nu W^-_\nu + \\
&\tfrac12 g^2 W^+_\mu W^-_\nu W^+_\mu W^-_\nu + g^2 c_w^2(Z^0_\mu W^+_\mu Z^0_\nu W^-_\nu - Z^0_\mu Z^0_\mu W^+_\nu W^-_\nu) + \\
&g^2 s_w^2(A_\mu W^+_\mu A_\nu W^-_\nu - A_\mu A_\mu W^+_\nu W^-_\nu) + g^2 s_w c_w[A_\mu Z^0_\nu(W^+_\mu W^-_\nu - \\
&W^+_\nu W^-_\mu) - 2A_\mu Z^0_\mu W^+_\nu W^-_\nu] - g\alpha[H^3 + H\phi^0\phi^0 + 2H\phi^+\phi^-] - \\
&\tfrac18 g^2\alpha_h[H^4+(\phi^0)^4+4(\phi^+\phi^-)^2+4(\phi^0)^2\phi^+\phi^-+4H^2\phi^+\phi^-+2(\phi^0)^2H^2] - \\
&gMW^+_\mu W^-_\mu H - \tfrac12 g\tfrac{M}{c_w^2}Z^0_\mu Z^0_\mu H - \tfrac12 ig[W^+_\mu(\phi^0\partial_\mu\phi^- - \phi^-\partial_\mu\phi^0) - \\
&W^-_\mu(\phi^0\partial_\mu\phi^+ - \phi^+\partial_\mu\phi^0)] + \tfrac12 g[W^+_\mu(H\partial_\mu\phi^- - \phi^-\partial_\mu H) - W^-_\mu(H\partial_\mu\phi^+ - \\
&\phi^+\partial_\mu H)] + \tfrac12 g\tfrac{1}{c_w}(Z^0_\mu(H\partial_\mu\phi^0 - \phi^0\partial_\mu H) - ig\tfrac{s_w^2}{c_w}MZ^0_\mu(W^+_\mu\phi^- - W^-_\mu\phi^+) + \\
&igs_w MA_\mu(W^+_\mu\phi^- - W^-_\mu\phi^+) - ig\tfrac{1-2c_w^2}{2c_w}Z^0_\mu(\phi^+\partial_\mu\phi^- - \phi^-\partial_\mu\phi^+) + \\
&igs_w A_\mu(\phi^+\partial_\mu\phi^- - \phi^-\partial_\mu\phi^+) - \tfrac14 g^2 W^+_\mu W^-_\mu[H^2+(\phi^0)^2+2\phi^+\phi^-] - \\
&\tfrac14 g^2\tfrac{1}{c_w^2}Z^0_\mu Z^0_\mu[H^2+(\phi^0)^2+2(2s_w^2-1)^2\phi^+\phi^-] - \tfrac12 g^2\tfrac{s_w^2}{c_w}Z^0_\mu\phi^0(W^+_\mu\phi^- + \\
&W^-_\mu\phi^+) - \tfrac12 ig^2\tfrac{s_w^2}{c_w}Z^0_\mu H(W^+_\mu\phi^- - W^-_\mu\phi^+) + \tfrac12 g^2 s_w A_\mu\phi^0(W^+_\mu\phi^- + \\
&W^-_\mu\phi^+) + \tfrac12 ig^2 s_w A_\mu H(W^+_\mu\phi^- - W^-_\mu\phi^+) - g^2\tfrac{s_w}{c_w}(2c_w^2-1)Z^0_\mu A_\mu\phi^+\phi^- - \\
&g^1 s_w^2 A_\mu A_\mu\phi^+\phi^- - \bar e^\lambda(\gamma\partial+m_e^\lambda)e^\lambda - \bar\nu^\lambda\gamma\partial\nu^\lambda - \bar u^\lambda_j(\gamma\partial+m_u^\lambda)u^\lambda_j - \bar d^\lambda_j(\gamma\partial+ \\
&m_d^\lambda)d^\lambda_j + igs_w A_\mu[-(\bar e^\lambda\gamma e^\lambda) + \tfrac23(\bar u^\lambda_j\gamma u^\lambda_j) - \tfrac13(\bar d^\lambda_j\gamma d^\lambda_j)] + \tfrac{ig}{4c_w}Z^0_\mu[(\bar\nu^\lambda\gamma^\mu(1+ \\
&\gamma^5)\nu^\lambda) + (\bar e^\lambda\gamma^\mu(4s_w^2-1-\gamma^5)e^\lambda) + (\bar u^\lambda_j\gamma^\mu(\tfrac43 s_w^2-1-\gamma^5)u^\lambda_j) + \\
&(\bar d^\lambda_j\gamma^\mu(1-\tfrac83 s_w^2-\gamma^5)d^\lambda_j)] + \tfrac{ig}{2\sqrt2}W^+_\mu[(\bar\nu^\lambda\gamma^\mu(1+\gamma^5)e^\lambda) + (\bar u^\lambda_j\gamma^\mu(1+ \\
&\gamma^5)C_{\lambda\kappa}d^\kappa_j)] + \tfrac{ig}{2\sqrt2}W^-_\mu[(\bar e^\lambda\gamma^\mu(1+\gamma^5)\nu^\lambda) + (\bar d^\kappa_j C^\dagger_{\lambda\kappa}\gamma^\mu(1+\gamma^5)u^\lambda_j)] + \\
&\tfrac{ig}{2\sqrt2}\tfrac{m_e^\lambda}{M}[-\phi^+(\bar\nu^\lambda(1-\gamma^5)e^\lambda) + \phi^-(\bar e^\lambda(1+\gamma^5)\nu^\lambda)] - \tfrac g2\tfrac{m_e^\lambda}{M}[H(\bar e^\lambda e^\lambda) + \\
&i\phi^0(\bar e^\lambda\gamma^5 e^\lambda)] + \tfrac{ig}{2M\sqrt2}\phi^+[-m_d^\kappa(\bar u^\lambda_j C_{\lambda\kappa}(1-\gamma^5)d^\kappa_j) + m_u^\lambda(\bar u^\lambda_j C_{\lambda\kappa}(1+ \\
&\gamma^5)d^\kappa_j] + \tfrac{ig}{2M\sqrt2}\phi^-[m_d^\lambda(\bar d^\lambda_j C^\dagger_{\lambda\kappa}(1+\gamma^5)u^\kappa_j) - m_u^\kappa(\bar d^\lambda_j C^\dagger_{\lambda\kappa}(1-\gamma^5)u^\kappa_j] - \\
&\tfrac g2\tfrac{m_u^\lambda}{M}H(\bar u^\lambda_j u^\lambda_j) - \tfrac g2\tfrac{m_d^\lambda}{M}H(\bar d^\lambda_j d^\lambda_j) + \tfrac{ig}2\tfrac{m_u^\lambda}{M}\phi^0(\bar u^\lambda_j\gamma^5 u^\lambda_j) - \tfrac{ig}2\tfrac{m_d^\lambda}{M}\phi^0(\bar d^\lambda_j\gamma^5 d^\lambda_j) + \\
&\bar X^+(\partial^2-M^2)X^+ + \bar X^-(\partial^2-M^2)X^- + \bar X^0(\partial^2-\tfrac{M^2}{c_w^2})X^0 + \bar Y\partial^2 Y + \\
&igc_w W^+_\mu(\partial_\mu\bar X^0 X^- - \partial_\mu\bar X^+ X^0) + igs_w W^+_\mu(\partial_\mu\bar Y X^- - \partial_\mu\bar X^+ Y) + \\
&igc_w W^-_\mu(\partial_\mu\bar X^- X^0 - \partial_\mu\bar X^0 X^+) + igs_w W^-_\mu(\partial_\mu\bar X^- Y - \partial_\mu\bar Y X^+) + \\
&igc_w Z^0_\mu(\partial_\mu\bar X^+ X^+ - \partial_\mu\bar X^- X^-) + igs_w A_\mu(\partial_\mu\bar X^+ X^+ - \partial_\mu\bar X^- X^-) - \\
&\tfrac12 gM[\bar X^+ X^+ H + \bar X^- X^- H + \tfrac{1}{c_w^2}\bar X^0 X^0 H] + \tfrac{1-2c_w^2}{2c_w}igM[\bar X^+ X^0\phi^+ - \\
&\bar X^- X^0\phi^-] + \tfrac{1}{2c_w}igM[\bar X^0 X^-\phi^+ - \bar X^0 X^+\phi^-] + igMs_w[\bar X^0 X^-\phi^+ - \\
&\bar X^0 X^+\phi^-] + \tfrac12 igM[\bar X^+ X^+\phi^0 - \bar X^- X^-\phi^0]
\end{aligned}
$$

**Figure 13**

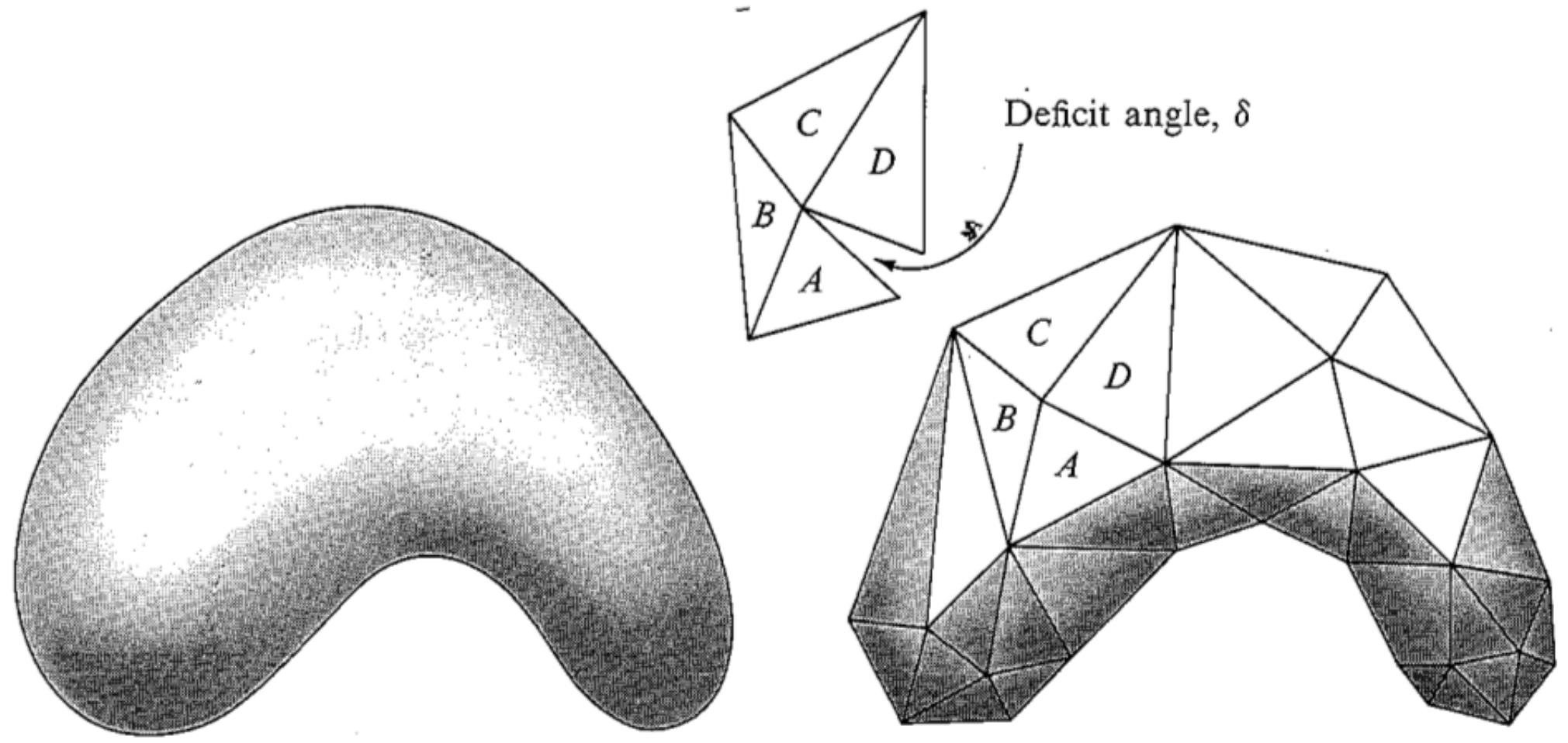

**Figure 42.1.**

A 2-geometry with continuously varying curvature can be approximated arbitrarily closely by a polyhedron built of triangles, provided only that the number of triangles is made sufficiently great and the size of each sufficiently small. The geometry in each triangle is Euclidean. The curvature of the surface shows up in the amount of deficit angle at each vertex (portion *ABCD* of polyhedron laid out above on a flat surface).

Reproduced from Misner, C.W., Thorne, K.S., Wheeler, J.A.: Gravitation. W.H. Freeman, San Francisco (1973), p. 1168.

**Figure 14**

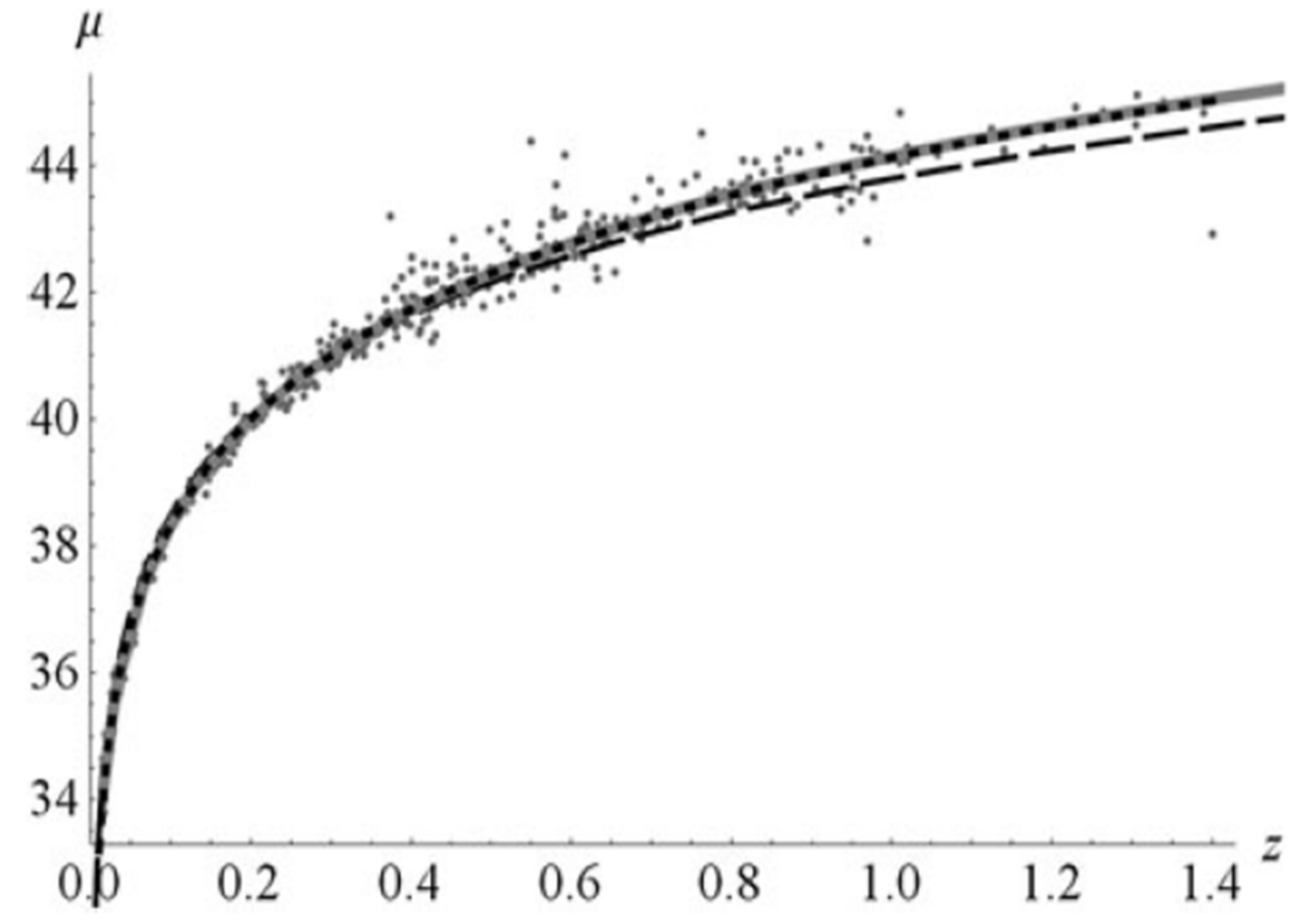

Plot of Union2 data along with the best fits for EdS (*dashed*), ΛCDM (*gray*), and MORC (*dotted*). The MORC curve is terminated at $z$ = 1.4 in this figure so that the ΛCDM curve is visible underneath.

---

[12] Maudlin, 2007

[13] Bell, J.S.: Speakable and Unspeakable. Cambridge University Press, Cambridge (1987), p 234

[14] Huggett, N., & Wüthrich, C.: Emergent Spacetime and Empirical (In)coherence. Studies in the History and Philosophy of Modern Physics 44, 276-285 (2013) http://arxiv.org/pdf/1206.6290.pdf

[15] Caravelli, F., & Markopoulou, F.: Disordered Locality and Lorentz Dispersion Relations: An Explicit Model of Quantum Foam (2012) http://arxiv.org/pdf/1201.3206v1.pdf;

Prescod-Weinstein, C., & Smolin, L.: Disordered Locality as an Explanation for the Dark Energy. Physical Review D 80, 063505 (2009) http://arxiv.org/pdf/0903.5303.pdf

[16] Weinstein, S.: Patterns in the Fabric of Nature. FQXi Essay Contest (Spring 2012) http://www.fqxi.org/community/forum/topic/1529

[17] Weinstein, 2012

[18] Silberstein *et al.*, 2013

[19] Carroll, S.: A Universe from Nothing? Discover Magazine Online (28 Apr 2012) http://blogs.discovermagazine.com/cosmicvariance/2012/04/28/a-universe-from-nothing/

[20] See examples in Oriti, D.: Disappearance and emergence of space and time in quantum gravity (2013) http://arxiv.org/abs/1302.2849

[21] Ambjorn, J., Goerlich, A., & Loll, R.: Quantum Gravity via Causal Dynamical Triangulations (2013) http://arxiv.org/abs/1302.2173

[22] Sorkin, R.: Relativity theory does not imply that the future already exists: a counterexample (2007) http://arxiv.org/pdf/gr-qc/0703098.pdf

[23] Rovelli, C.: 'Localization' in quantum field theory: How much of QFT is compatible with what we know about space-time?: In: Cao, T. (ed.) Conceptual Foundations of Quantum Field Theory, pp 207-232, Cambridge University Press, Cambridge (1999), p 227

[24] Weyl, H.: Space, Time, Matter. Dover Publications, New York (1952)

[25] Healey, R.: Gauging What's Real: The Conceptual Foundations of Gauge Theories. Oxford University Press, Oxford (2007), p 47

---

[26] Ladyman, J.: Structural Realism: In: Zalta, E.N. (ed.) The Stanford Encyclopedia of Philosophy (Summer 2009 Edition) http://plato.stanford.edu/archives/sum2009/entries/structural-realism;

See also Ladyman, J., Ross, D., Sturrett, D., & Collier, J.: Everything Must Go: Metaphysics Naturalized. Oxford University Press, Oxford (2007)

[27] Ladyman, 2009

[28] Ladyman *et al*., 2007;

Rickles, D.: Time Observables and Structure: In: Landry, E., & Rickles, D. (ed.) Structural Realism: Structure, Object, and Causality, pp 135-148, Springer, New York (2012a), p 144

[29] Esfeld, M.: Ontic structural realism and the interpretation of quantum mechanics. European Journal For Philosophy of Science 3(1), 19-32 (2013)

[30] Rickles, D., & Bloom, J.: Things Ain't What They Used to Be. FQXi Essay Contest (Spring 2012b) http://www.fqxi.org/community/forum/topic/1537

[31] Rickles & Bloom, 2012b

[32] Rickles & Bloom, 2012b

[33] Silberstein *et al*., 2008

[34] Geroch, R.: General Relativity from A to B. University of Chicago Press, Chicago (1978), pp 20-21

[35] Henneaux, M., & Teitelboim, C.: Quantization of Gauge Systems. Princeton University Press, Princeton (1992), p 106

[36] Kiefer, C.: Time in quantum gravity: In: Callender, C. (ed.) The Oxford Handbook of Philosophy of Time, Oxford University Press, Oxford (2011), p 667

[37] Wharton, K.: Quantum States as Ordinary Information. Information 5, 190-208 (2014)

[38] Rickles, 2012a, p 141

[39] Misner, C.W., Thorne, K.S., & Wheeler, J.A.: Gravitation. W.H. Freeman, San Francisco (1973), p 364

[40] Misner *et al*., 1973, p 369;

Wise, D.K.: p-Form electromagnetism on discrete spacetimes. Classical and Quantum Gravity 23, 5129-5176 (2006)

---

[41] Wise, 2006

[42] Zee, A.: Quantum Field Theory in a Nutshell. Princeton University Press, Princeton (2003), p 172

[43] Zee, 2003, p 173

[44] Shankar, R.: Principles of Quantum Mechanics, 2nd Ed. Plenum Press, New York (1994), p 226

[45] Zee, 2003, p 22

[46] See for example: http://www.phy.ohiou.edu/~elster/lectures/relqm_17.pdf .

[47] Bach, R., Pope, D., Sy-Hwang, L., & Batelaan, H.: Controlled double-slit electron diffraction. New Journal of Physics 15, 033018 (2013)

[48] Rothe, H.J.: Lattice Gauge Theories, An Introduction. World Scientific, Singapore (1992), p 72

[49] Padmanabhan, T.: From Gravitons to Gravity: Myths and Reality (2004) http://arxiv.org/pdf/gr-qc/0409089v1.pdf

[50] Zee, 2003, p 424

[51] ‘t Hooft, G.: The Conceptual Basis of Quantum Field Theory: In: Butterfield, J., Earman, J. (eds.) Philosophy of Physics Part A, pp 661-729, North-Holland, Amsterdam (2007)

[52] Rothe, 1992, p 40

[53] Frisch, H.J.: Pattern recognition at the Fermilab collider and Superconducting Supercollider. Proceedings of the National Academy of Sciences 90, 9754-9757 (November 1993)

[54] Mott, N.: The wave mechanics of alpha-ray tracks. Proceedings of the Royal Society A126, 79-84 (1929)

[55] Mott, 1929

[56] Colosi, D., & Rovelli, R.: What is a particle? Classical and Quantum Gravity 26, 025002 (2009) http://arxiv.org/abs/gr-qc/0409054

[57] Colosi & Rovelli, 2009

[58] Stuckey *et al.*, 2012a

---

[59] Amanullah, R., Lidman, C., Rubin, D., Aldering, G., Astier, P., Barbary, K., Burns, M.S., Conley, A., Dawson, K.S., Deustua, S.E., Doi, M., Fabbro, S., Faccioli, L., Fakhouri, H.K., Folatelli, G. Fruchter, A.S., Furusawa, H., Garavini, G., Goldhaber, G., Goobar, A., Groom, D.E., Hook, I., Howell, D.A., Kashikawa, N., Kim, A.G., Knop, R.A., Kowalski, M., Linder, E., Meyers, J., Morokuma, T., Nobili, S., Nordin, J., Nugent, P.E., Ostman, L., Pain, R., Panagia, N., Perlmutter, S., Raux, J., Ruiz-Lapuente, P., Spadafora, A.L., Strovink, M., Suzuki, N., Wang, L., Wood-Vasey, W.M., & Yasuda, N. (The Supernova Cosmology Project): Spectra and light curves of six type Ia supernovae at $0.511 < z < 1.12$ and the Union2 compilation. Astrophysics Journal 716, 712–738 (2010) http://arxiv.org/abs/1004.1711

[60] Komatsu, E., Smith, K.M., Dunkley, J., Bennett, C.L., Gold, B., Hinshaw, G., Jarosik, N., Larson, D., Nolta,M.R., Page, L., Spergel, D.N., Halpern, N., Hill, R.S., Kogut, A., Limon, M.,Meyer, S.S., Odegard, N., Tucker, G.S., Weiland, J.L., Wollack, E., & Wright, E.L.: Seven-Year Wilkinson Microwave Anisotropy Probe (WMAP) Observations: Cosmological Interpretation. (2010) http://arxiv.org/abs/1001.4538

[61] Riess, A.G., Macri, L., Casertano, S., Lampeitl, H., Ferguson, H.C., Filippenko, A.V., Jha, S.W., Li, W., & Chornock, R.: A 3% solution: determination of the Hubble constant with the Hubble space telescope and wide field camera 3. Astrophysical Journal 730, 119 (2011)

[62] Stuckey *et al*., 2012b

[63] Carroll, 2012